\documentclass[10pt,journal, compsoc]{IEEEtran}
\usepackage{amsmath, amsfonts, amssymb, cite, epsfig, mathtools,adjustbox}
\usepackage{algorithm, algorithmic, amsbsy,bm,bbm,amsthm}
\usepackage{xcolor}
\usepackage{adjustbox,multirow, booktabs, hhline,tabularx, multirow}
\usepackage{subcaption}
\usepackage{dblfloatfix}
\usepackage{threeparttable}

\def\hlb{\color{black}}
\def\hlr{\color{black}}

\newtheorem{theorem}{Theorem}
\newtheorem{lemma}{Lemma}

\newtheorem{corollary}{Corollary}
\newtheorem{assumption}{Assumption}

\usepackage{bm, bbm}

\makeatletter
\newcommand{\multiline}[1]{%
  \begin{tabularx}{\dimexpr\linewidth-\ALG@thistlm}[t]{@{}X@{}}
    #1
  \end{tabularx}
}

\usepackage{epstopdf}
\graphicspath{{figures/}}

\def \FigWidthSmaller{0.34\textwidth}
\def \FigWidthSmall{0.38\textwidth}

\def \w{\mathbf{w}}

\def\1{\mathbf 1}

\def\E{\mathbb E}
\def\N{\mathcal N}
\def\K{\mathcal K}
\def\M{\mathcal M}

\begin{document}
%
\title{On the Convergence of Multi-Server Federated Learning with Overlapping Area}
%
%
%
%

\author{Zhe~Qu,~\IEEEmembership{Student~Member,~IEEE,}
        Xingyu~Li,~\IEEEmembership{Student~Member,~IEEE,}
        Jie~Xu,~\IEEEmembership{Senior~Member,~IEEE,}
        Bo~Tang,~\IEEEmembership{Senior~Member,~IEEE,}
        Zhuo~Lu,~\IEEEmembership{Senior~Member,~IEEE,}
        and~Yao~Liu,~\IEEEmembership{Senior~Member,~IEEE}
\IEEEcompsocitemizethanks{\IEEEcompsocthanksitem Z. Qu and Z. Lu are with the Department of Electrical Engineering, University of South Florida, Tampa, 33620, USA. E-mail: \{zhequ, zhuolu\}usf.edu.
\IEEEcompsocthanksitem X. Li and B. Tang are with the Department of Electrical and Computer Engineering, Mississippi State University, Starkville, MS, 39759, USA. E-mail: \{xl292@, tang@ece.\}msstate.edu.
\IEEEcompsocthanksitem J. Xu is with the Department of Electrical and Computer Engineering, University of Miami, Coral Gables, FL, 33146, USA. E-mail: jiexu@miami.edu.
\IEEEcompsocthanksitem Y. Liu is with the Department of Computer Science and Engineering, University of South Florida, Tampa, FL, 33620, USA. E-mail: yliu@cse.usf.edu.
\IEEEcompsocthanksitem Z. Qu and X. Li are co-first author.}}

\IEEEtitleabstractindextext{%
\begin{abstract}
Multi-server Federated learning (FL) has been considered as a promising solution to address the limited communication resource problem of single-server FL. We consider a typical multi-server FL architecture, where the coverage areas of regional servers may overlap. The key point of this architecture is that the clients located in the overlapping areas update their local models based on the average model of all accessible regional models, which enables indirect model sharing among different regional servers. Due to the complicated network topology, the convergence analysis is much more challenging than single-server FL. In this paper, we firstly propose a novel MS-FedAvg algorithm for this multi-server FL architecture and analyze its convergence on non-iid datasets for general non-convex settings. Since the number of clients located in each regional server is much less than single-server FL, the bandwidth of each client should be large enough to successfully communicate training models with the server, which indicates that full client participation can work in multi-server FL. Also, we provide the convergence analysis of the partial client participation scheme and develop a new biased partial participation strategy to further accelerate convergence. Our results indicate that the convergence results highly depend on the ratio of the number of clients in each area type to the total number of clients in all three strategies. The extensive experiments show remarkable performance and support our theoretical results. 
\end{abstract}

}

\maketitle

\IEEEdisplaynontitleabstractindextext

%
\IEEEpeerreviewmaketitle

\IEEEraisesectionheading{\section{Introduction}\label{sec:introduction}}

With the explosive growth in the numbers of mobile phones and Internet of Things (IoT) devices, a tremendous amount of data today is being generated at the network edge in a distributed manner. Sending this data to the cloud for processing not only puts a huge burden on the network but also raises serious data privacy concerns. Federated Learning (FL) \cite{konevcny2016federated, mcmahan2017communication, stich2018local} recently emerged as a distributed Machine Learning (ML) architecture that keeps all the training data on individual clients, thereby protecting client data privacy and mitigating network congestion.

In its most common form, FL is an iterative process where in each communication round, clients train local ML models using their local training datasets based on the current global ML model, and then the server aggregates the local models uploaded by the clients to update the global model. Because FL is trained on distributed datasets and often involves many communication rounds of model data exchange between the clients and the server, improving the communication efficiency between clients and central server \cite{zhao2020semi, haddadpour2021federated} and handling heterogeneous local dataset distribution in each client \cite{li2019convergence, yang2021achieving} are two biggest challenges of FL and have received a large amount of research attention.

Although promising progresses have been made, existing FL architectures and algorithms dominantly focus on the single-server system. Most FL studies consider that clients should download and upload the learning models with central server repeatedly in each communication round. This communication strategy may suffer a large communication delay in large-scale FL systems where many clients may be far away from the server \cite{mao2017survey, nguyen2019joint}. This large delay between the server and the clients directly prolongs the learning time of the existing single-server-based FL system, especially when the server is placed on the cloud. As increasingly many applications are delay-sensitive, e.g., autonomous driving and wearable health monitoring, new FL architectures that involve multiple servers have been proposed to reduce the communication latency between the server and the clients. 

\begin{figure*}[t!]
    \centering
    \includegraphics[width=\textwidth, height=0.3\textwidth]{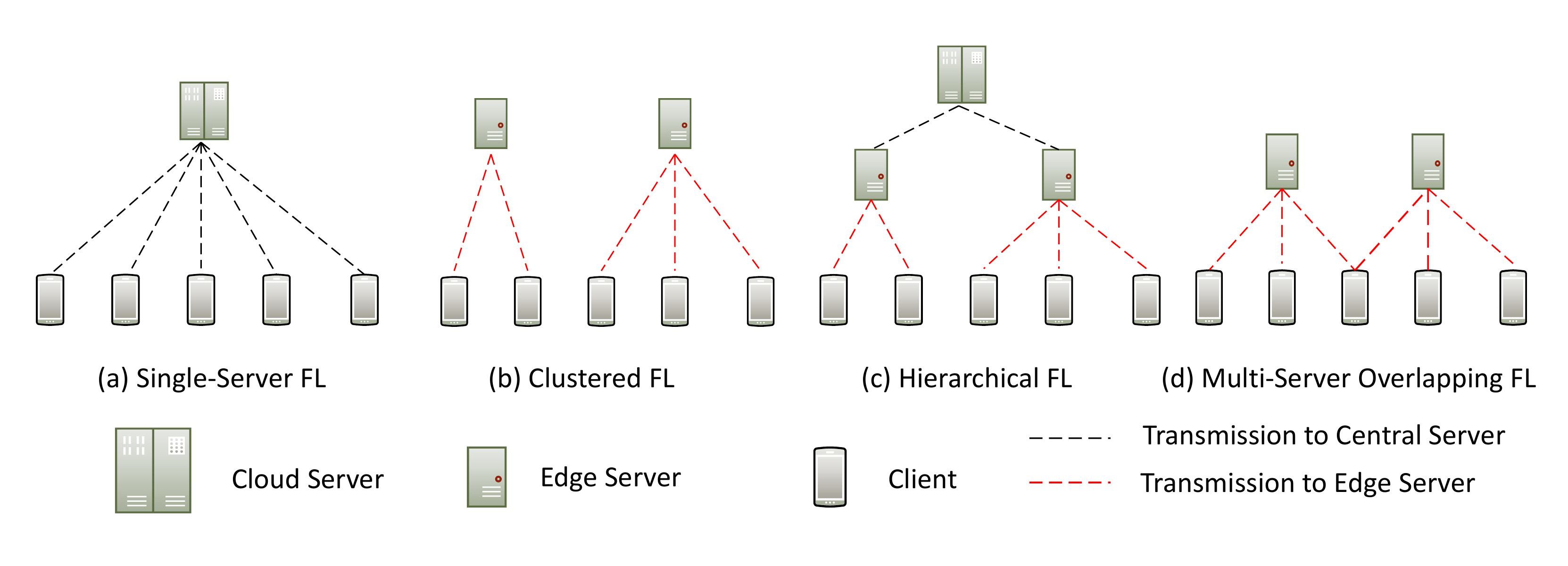}
    \caption{Description of four different FL network architectures: (a) {\hlr Single-server FL}; (b) Clustered FL; (c) Hierarchical FL and (d) {\hlr Multi-server overlapping FL.}}
    \label{Fig:differentFlsetting}
\end{figure*}

In order to reduce the communication latency of FL, there are two main multiple servers FL approaches: (1) Hierarchical FL (HFL) \cite{lim2020federated, liu2020client, wang2020local} introduces a hierarchical structure for model training where multiple edge servers are used to collect and aggregate local model updates from clients in their respective service areas and then send the aggregated result to the cloud server for final aggregation. However, since the models exchange between the edge servers and the cloud server is still required, HFL can still result in a long training delay when the propagation latency between the edge servers and the cloud server is large. (2) Clustered FL (CFL) \cite{xie2020multi, lee2020tornadoaggregate, duan2020self} divides clients into different clusters, and trains a separate ML model for each cluster. However, re-clustering computation may be required in many communication rounds, thereby significantly increasing the training complexity and time. In addition, some existing studies ignore the physical network connectivity constraints -- a client may connect to only a subset of servers. 

In \cite{han2021fedmes}, a new FL architecture utilizing multiple servers is studied, which exploits the realistic deployment of 5G-and-beyond networks where a client can be located in the overlapping coverage areas of multiple servers. The network architectures of single-server FL, HFL, clustered FL and our proposed multi-server FL are shown in Fig.~1. The key idea is that clients download multiple models from all the edge servers they can access and train their local models based on the average of these models. Such an architecture has two main advantages. First, by performing model averaging on the client side, each server indirectly accesses the trained local models of clients not in its coverage while incurring a small model upload and download delay. Specifically, the broadcasting technique will not increase the communication burden. Second, instead of training multiple local models based on multiple downloaded models, each client only trains a single local model based on the average of the downloaded models, thereby avoiding extra computation and communication cost. {\hlb Since the clients in overlapping areas should tackle multiple training models at the same time, the extra computation only comes from the averaging calculation, which is small compared to the local training process, and it can be negligible.} Although \cite{han2021fedmes} developed an algorithm for this architecture and empirically validated its effectiveness, they only proposed the strongly-convex loss function, which is very restricted, since most learning models are non-convex, e.g., neural network. In addition, the convergence results cannot show the impact on overlapping areas. In this paper, we improve upon this architecture, propose a new algorithm with two-sided learning rates, and provide theoretical convergence analysis of the more general non-convex loss function. In summary, we highlight our main contributions as follows:

1) We develop a novel MS-FedAvg algorithm on this multi-server FL architecture, based on the two-sided learning rates FedAvg.

2) We study the convergence in the coverage area, where we call region, of each server. For non-convex loss functions and non-iid datasets, we provide convergence analysis for full and unbiased partial client participation strategies, respectively. Our results are better than the existing multi-server FL algorithms and also reveal how the overlapping coverage affects the convergence in each region.

3) To further improve the convergence speed of MS-FedAvg, we develop a biased partial client participation strategy where clients may not be selected proportionally to the number of clients in different coverage areas. Our analysis shows that the degree of bias results in a trade-off between convergence rate and accuracy. 

4) We conduct extensive experiments on multiple datasets under different multi-server FL network architectures and hyper-parameters. The experimental results show that our MS-FedAvg algorithm outperforms the compared benchmarks from accuracy and convergence perspectives.

The preliminary of FL is presented in Section~\ref{Sec:System}. In Section~\ref{Sec:msfedavg}, we develop the MS-FedAvg algorithm for our proposed multi-server FL architecture. Section~\ref{Sec:ConvergenceMSfedavg} analyzes the convergence rate of MS-FedAvg including full, unbiased partial and biased partial client participation strategies. The discussion of MS-FedAvg is presented in Section~\ref{Sec:Discussion}. {\hlb In Section~\ref{Sec:latency}, we present the transmission latency of different FL architecture.} Experimental results are shown in Section~\ref{Sec:experiment}. Section~\ref{Sec:Related} overviews the related works, followed by the conclusion in Section~\ref{Sec:Conclusion}.

\section{Preliminary of FL}\label{Sec:System}
We consider a FL network including a number of $N$ clients, indexed by $\mathcal{N} = \{1, ..., N\}$ and one central server/aggregator, where each client $i \in \mathcal{N}$ has its own local dataset with the data distribution $D_i$. FL aims to solve the following risk minimization problem:
\begin{small}
\begin{align}\label{Eq:Fedavg}
    \min_{\w} \left\{ f(\w) \triangleq \frac{1}{N}\sum_{i=1}^{N}F_{i}(\w)\right\},
\end{align}
\end{small}
where $F_i(\w) \triangleq \E_{\xi \sim D_i}[F_i (\w ,\xi )]$ is the local loss function. FedAvg \cite{mcmahan2017communication}, a seminal FL algorithm, works in an iterative manner as follows:

1) In each communication round $t$, each client $i$ downloads the current global model $\w^t$ from the server and sets its initial local model as the current global model, i.e., $\w_i^{t,0} = \w^t$. 

2) Each client runs $E$ steps of Stochastic Gradient Descent (SGD) as follows:
\begin{small}
\begin{equation}\label{Eq:SGD}
    \w^{t,e+1}_{i} = \w_i^{t, e} - \eta_l \nabla F_{i} (\w_i^{t,e}), \forall e = 0,\dots, E-1 ,
\end{equation}
\end{small}
where $\eta$ is the learning rate of local training. Client $i$'s updated model after these $E$ steps can be written as $\w_i^{t+1} = \w_i^{t,E}$. 

3) Each client uploads the updated model $\w_i^{t+1}$ to the server, which computes a simple aggregation $\w^{t+1} = \frac{1}{N}\sum_{i=1}^{N}\w_i^{t+1}$. 

Due to the computation and bandwidth limitation, full client participation is often not practical. Hence, the more realistic FL strategy is that the server can select a subset of $K$ clients, indexed by $\mathcal{K}^t \subseteq \mathcal{N}$, to participate in FL in communication round $t$, and the global model is computed according to $\w^{t+1} = \frac{1}{K}\sum_{i\in \mathcal{K}^t}\w_{i}^{t+1}$. This is known as partial client participation strategy \cite{li2019convergence, yang2021achieving,  karimireddy2020scaffold}.

\section{Multi-Server FedAvg (MS-FedAvg)}\label{Sec:msfedavg}
A single-server FL system may incur a large delay if clients are distributed in the network, some of which may be far away from the server. Developing a multi-server FL network architecture is a potential way to address this problem, e.g., Hierarchical FL (HFL) \cite{lim2020federated, liu2020client, wang2020local} and Clustered FL (CFL) \cite{xie2020multi, lee2020tornadoaggregate, duan2020self}. While HFL requires the models on edge server should be aggregated on the global server every several communication rounds, it also incurs the extra transmission delay. CFL needs to re-cluster the clients every communication round based on a specific rule, e.g., model similarity or location, it is difficult to avoid the clients are far away from the edge server which should be clustered. 

Since all the existing multi-server FL network architectures cannot directly leverage solve the large delay problem, we consider a multi-server FL architecture as in \cite{han2021fedmes}, where multiple \textbf{regional servers} are distributed in close proximity to the clients. Let $\mathcal{M} = \{1, ..., M\}$ be the set of regional servers and each regional server $m$ covers a subset of client $\mathcal{N}_m \subseteq \mathcal{N}$ with $|\mathcal{N}_m| = N_m$. For convenience, we call $\mathcal{N}_m$ \textbf{region} $m$. It is worth noting that a client may locate in multiple regions, because the coverage areas of the servers may overlap. 

\begin{figure}[t!]
    \centering
    \includegraphics[width=\FigWidthSmall, height=0.24\textwidth]{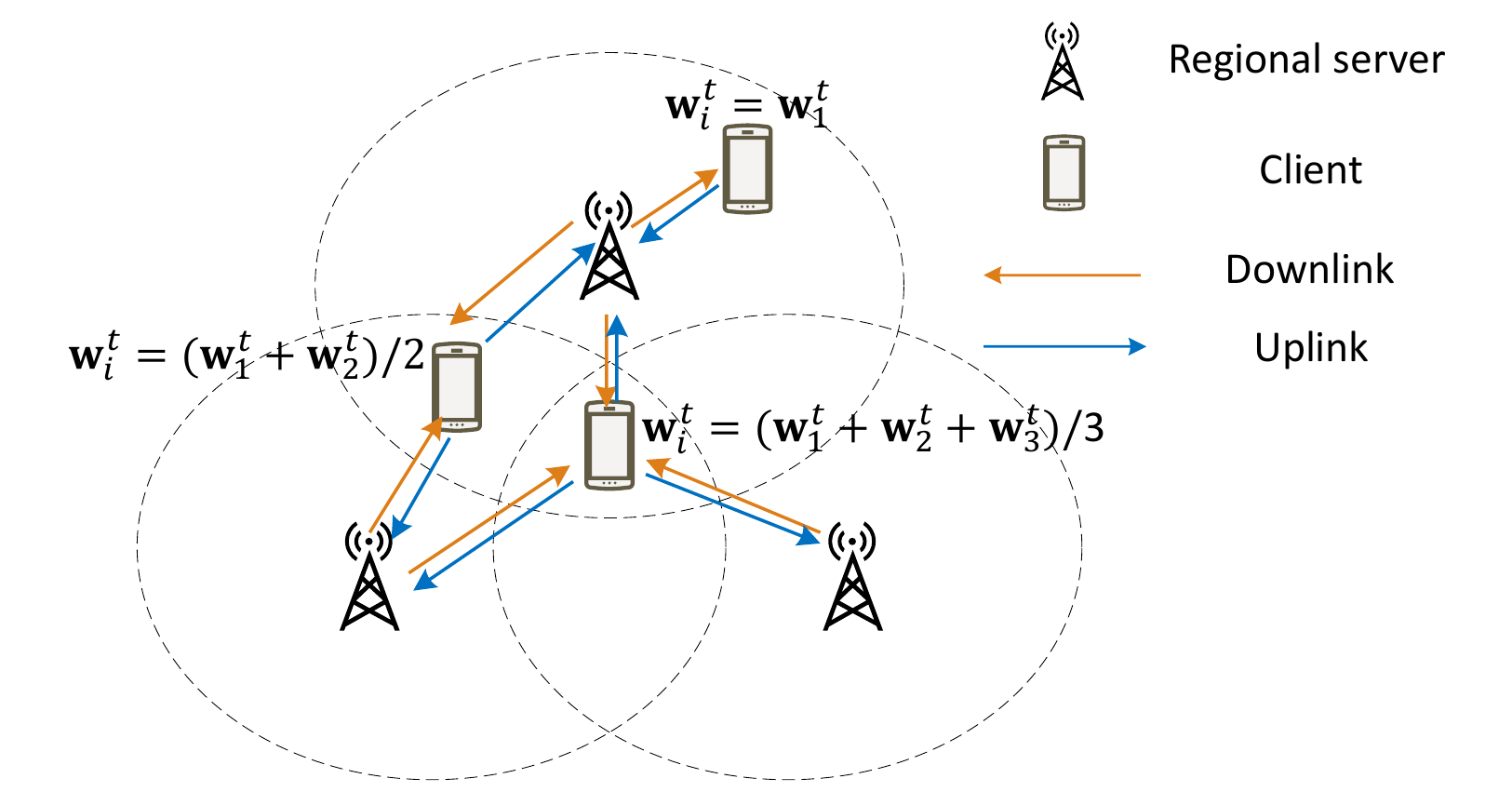}
    \caption{Description of MS-FedAvg: $\w_i^t$ is local training model on clients, and $\w_1^t, \w_2^t, \w_3^t$ are regional learning models on servers.}
    \label{Fig:DescriptionMSFL}
\end{figure}

In MS-FedAvg, each regional server trains a \textbf{regional model} using clients in its region, and a client updates its local model based on all regional models that it can access, where the architecture with three regional servers is shown in Figure~\ref{Fig:DescriptionMSFL}. Different from HFL, the regional models are not aggregated until the final round to generate a global model. Let $\mathcal{M}_i \subseteq \mathcal{M}$ be the set of regional servers that client $i$ can communicate where $M_i = |\mathcal{M}_i|$, and $\M_i^t \subseteq \M$ be the set of servers that client $i$ is sampled in communication round $t$. 

At the beginning of a communication round $t$, any client $i$ downloads the current regional models $\w^t_m, \forall m \in \mathcal{M}_i$ from all the $M_i$ servers, and averages the downloaded regional models to be the initial local model in the current round, i.e., $\w_i^{t, 0} = \frac{1}{M_i}\sum_{m \in \mathcal{M}_i}\w_{m}^t$. Then, each client updates its local model using SGD for $E$ local training epochs to obtain the local model $\w_i^{t+1}$ by \eqref{Eq:SGD}, and uploads it to the servers in $\mathcal{M}^t_i$. Each server $m$ then updates the regional model according to $\w^{t+1}_m = \frac{\eta_g}{N_m}\sum_{i \in \mathcal{N}^t_m}\w^{t+1}_i$, where $\eta_g$ is the regional learning rate. After a sufficient number of $T$ communication rounds, the global model is finally obtained by averaging over the converged regional models, i.e., $\w = \frac{1}{M}\sum_{m\in\mathcal{M}} \w^T_m$.

\begin{algorithm}[t!]
	\caption{MS-FedAvg algorithm.}
	\label{ALG:MSFL}
	\begin{algorithmic}[1]
		\STATE {\bf Input:} Initialize model $\w_m^0$ to each server $m$.
		\STATE {\bf Output:} Final global model $\w$.
		\STATE Set $\w_i^0 = \w^0$ for all clients $i = 1,2, \dots, N$;
		\FOR {$t = 0, T-1$}
		\FOR {Server $m = 1,\dots,M$}
		\FOR {$i = 1,\dots, N_m$ in parallel}
		\IF {Client $i$ is in non-overlapping area}
		\STATE $\w_{i,0}^{t} = \w_m^t$; 
		\ELSE 
		\STATE $\w_i^{t, 0} = \frac{1}{M_i}\sum_{m \in \mathcal{M}_i}\w_{m}^t$; 
		\ENDIF
		\STATE Computes $E$ local training epochs from Eq.~\eqref{Eq:SGD} and uploads $\w_{i,E}^{t}$ to the connecting server(s);
		\STATE Regional model: $\w_{m}^{t+1}=\frac{\eta_g}{N_m}\sum_{i\in\mathcal{N}_m^t}\w_{i,E}^t$;
		\ENDFOR
		\ENDFOR
		\ENDFOR
		\STATE Global model: $\w = \frac{1}{M}\sum_{m\in\mathcal{M}} \w^T_m$.
	\end{algorithmic}
\end{algorithm}

The pseudo-code of MS-FedAvg is given in Algorithm~\ref{ALG:MSFL}. {\hlb Compared to the single-server FL \cite{konevcny2016federated, mcmahan2017communication, li2019convergence, karimireddy2020scaffold}, a unique feature of MS-FedAvg is that clients in overlapping areas receive and average multiple regional models to be the initial model for local training in each communication round (Line 10). Together with the model averaging at the servers, this two-sided model averaging process allows the servers to indirectly access the local models of clients outside their regions instead of combining local model updates from the clients in overlapping areas (Line 13), which bridges the regional model sharing, thereby fully utilizing all clients' data in the network.} {\hlr Specifically, optimizing the placement of the limited number of regional servers in the current mobile computing system to maximize the total coverage is considered rather important. For example, some popular ES placement algorithms \cite{lai2018optimal, cui2020robustness} have shown that one overlapping area at most includes four regional servers. Therefore, we consider that the additional transmission latency of the overlapping areas clients mentioned in the paper can be very small and negligible.} Although the clients located in the overlapping area, we can leverage the broadcast technique that cannot incur high burden of communication. 

{\hlb Since we consider that the regional server is similar to the edge server, where the coverage is very restricted \cite{chen2019budget}, the number of clients in the region of each regional server are much fewer than single-server FL. Therefore, the communication burden is less than single-server FL due to the shorter distance and more stable connection, and hence full client participation should work well in our proposed multi-server FL architecture.} In this paper, we only consider the location of clients is fixed for proving the convergence results of MS-FedAvg. {\hlb The static scenario can be considered as the hospital data \cite{xu2021federated} or environmental monitoring sensors \cite{liu2022data}, where the clients cannot move and only connect to the corresponding regional server(s). As such, our MS-FedAvg can improve this scenario efficiently.} If the clients move randomly, each regional model can be assumed as an individual FL model approaching by FedAvg, which might degrade the training performance. In Section~\ref{Sec:experiment}, we propose the experimental results of the movement scenario under our multi-server FL architecture. {\hlr More specifically, \cite{kong2021federated, ouyang2021clusterfl} developed the FL-based license plate recognition and human activity recognition algorithm. Firstly, the recognition results can be quickly obtained due to the less transmission latency. On the other hand, when clients come into overlapping areas, multiple monitors can acquire more information and recognize them more accurately.}

Because the regional models are averaged only in the final communication round, a significant amount of communication and computation cost among the regional servers can be saved. However, the model averaging at the client side also introduces an obvious difference compared to single-server FedAvg: the initial models of the clients, even for those in the same region, for local training in each communication round can be different depending on the their specific locations. Since we consider that the regional server is edge server and the coverage area is very restricted, the number of clients in the region of each regional server are much fewer than single-server FL. Therefore, the communication burden is less than single-server FL, and hence full client participation should work well in our proposed multi-server FL. {\hlb In order to clarify the reducing of transmission latency compared to other FL architectures, we will show the detailed quantification in Section~\ref{Sec:latency}.}

In this paper, we also consider the partial client participation strategy, which is a more realistic strategy for single-server FL \cite{mcmahan2017communication, li2019convergence, karimireddy2020scaffold, yang2021achieving}. More specifically, at the beginning of a communication round $t$, each server $m$ randomly samples a subset of clients $\mathcal{K}^t_m \subseteq \mathcal{N}_m$ in its region to participate in the current round's training, with $K_m = |\mathcal{K}^t_m|$. Because a client may be in multiple regions, it may be sampled by multiple servers, which brings more challenges for convergence analysis compared to single-server FL. We also provide the convergence analysis of partial client participation strategy of MS-FedAvg.

\section{Convergence Analysis of MS-FedAvg}\label{Sec:ConvergenceMSfedavg}
In this section, we focus on a representative region $\mathcal{N}_m$ and study the convergence of its regional model. Let $f_m(\w) = \frac{1}{N_m}\sum_{i\in\mathcal{N}_m} F_i(\w)$ be the objectives of region $m$. As discussed in the last section, the main difficulty of the convergence analysis lies in the heterogeneous initial models of clients in the region in each communication round. The convergence analysis encompasses non-iid datasets for general non-convex loss settings under both full and partial client participation strategies. Besides the existing random participation, i.e., unbiased client participation \cite{li2018federated, li2019convergence, li2020federated, karimireddy2020scaffold, stich2018local}, we also propose a new biased client participation strategy for our propose MS-FedAvg algorithm. To propose convergence results of MS-FedAvg, we first state some useful assumptions in this paper as follows:

\begin{assumption}\label{Ass:Lsmooth}
(Lipschitz Gradient) $\forall i \in \mathcal{N}_m$, $F_i$ is $L$-smooth, i.e., for all $\mathbf{v}$ and $\mathbf{w}$, 
\begin{small}
\begin{equation}
    F_i (\mathbf{v})\leq F_i (\w) + (\mathbf{v} - \w)^T \nabla F_i (\w) + \frac{L}{2}\|\mathbf{v} - \w\|^2_2 .\nonumber
\end{equation}
\end{small}
\end{assumption}

\begin{assumption}\label{Ass:Unbiased}
(Unbiased Local Gradient Estimator) Let $\xi_i$ be a random local data sample on client $i$. $\forall i \in \mathcal{N}_m$ and $\forall \w$, the local gradient estimator is unbiased, i.e., $\E [\nabla F_i(\w , \xi_i)]=\nabla F_i(\w)$, where the expectation is taken over all the local datasets samples. 
\end{assumption}

\begin{assumption}\label{Ass:variancegradient}
(Bounded Local Variance) $\forall i\in \mathcal{N}_m$ and $\forall \w$, the variance of local gradient estimator of any regional server $m$ can be upper-bounded by a constant $\sigma_m$, i.e., 
\begin{small}
\begin{equation}
\E\|\nabla F_i(\w, \xi_i) - \nabla F_i (\w)\|^2 \leq \sigma_m^2.\nonumber
\end{equation}
\end{small}
\end{assumption}

Assumptions~\ref{Ass:Lsmooth}-\ref{Ass:variancegradient} are fairly standard in existing FL works \cite{li2019convergence, yu2019parallel, qu2020federated}. For the following assumption, we need to introduce the notion of the \textit{type} of a client. Even for clients in the same region, they differ in terms of the subset of servers they may access since they may be in different overlapping coverage areas. Thus, we say that two clients have the same type if they can access the same set of servers. Formally, we define the client type $\theta \subseteq 2^\mathcal{M}$ to be the subset of servers that it can access. Let $\mathcal{K}^t_{m, \theta}$ be the set of clients of type $\theta$ that is sampled in region $m$ in round $t$, and let ${K}_{m, \theta}^t=|\mathcal{K}^t_{m, \theta}|$. Clearly, for all clients in region $m$, $m$ must be an element of their types. Moreover, if two regions $m$ and $m'$ do not overlap, then there must be no client whose type contains both $m$ and $m'$. 

\begin{assumption}\label{Ass:regionalbound}
(Bounded Regional Variance) For any client $i$ of type $\theta$ in region $m$ and for any round $t$, the gradient difference of its local loss function at $\w_i^{t+1}$ and the regional loss function at $\w_m^{t+1}$ is upper-bounded, i.e., 
\begin{equation}
\|\nabla F_i(\w^{t+1}_i) - \nabla f_m(\w^{t+1}_m)\|^2 \leq \alpha^2_{m,\theta}.\nonumber
\end{equation}
\end{assumption}
Assumption 4 states that clients of different types have different impacts on the gradient of the regional loss function at the end of each round. This impact is a joint result of the non-iid local datasets and different initial model at the beginning of the training round due to different coverage areas, i.e., types, which is different from the single-server FL \cite{yang2021achieving, reddi2020adaptive}. It is worth noting that in this paper we do not assume to bound gradient descent \cite{li2019convergence, han2021fedmes}, i.e., $\|\nabla f(\w)\|^2 \leq G^2$, where it is a loose assumption.

Before analyzing the convergence results of MS-FedAvg algorithm, we first propose the key lemma for both full and partial client participation strategies, which aims to propose the upper-bound of client drift for every regional model. 
\begin{lemma}\label{Lemma:combinedvariance}
For any $\eta_l < \frac{1}{\sqrt{30}LE}$, we have the following results:
\begin{small}
\begin{equation}
\begin{split}
    &\frac{1}{N_{m,\theta}}\sum_{i\in\N_{m, \theta}}\mathbb{E} \|\w^{t,e}_i - \w_m^t \|^2 \\
    & \leq 5E\eta_l^2 \bigg(\sigma_m^2 + \frac{6EN_{m,\theta}}{N}\alpha^2_{m,\theta}\bigg)+ 30E^2\eta_l^2 \|\nabla f (\w^t_m )\|^2 .\nonumber
\end{split}
\end{equation}
\end{small}
\end{lemma}
\noindent\textit{Proof.} The proof is shown in Appendix~A.\hfill~$\Box$

\subsection{Convergence Analysis for Full Client Participation}\label{SubSec:convergence}
For the full client participation strategy of MS-FedAvg, we have the following convergence result:
\begin{theorem}\label{The:nonconvexfull}
Let assumptions~\ref{Ass:Lsmooth}-\ref{Ass:regionalbound} hold and $L, \sigma_{m}^2, \alpha_{m, \theta}^2$ be defined therein. With full client participation strategy, if we choose the learning rate $\eta_l \leq \min\{\frac{1}{\sqrt{30}LE},\frac{1}{LE\eta_g}\}$. The convergence result is given as follows:
\begin{small}
\begin{equation}
    \min_{t \in [T]} \mathbb{E}\|\nabla f (\w^t )\|^2 \leq \frac{f^0 - f^* }{cMET\eta_g \eta_l} + \Psi, \nonumber
\end{equation}
\end{small}
where $c$ is a constant, $f^0 \triangleq f_m (\w^0 )$, $f^* \triangleq f (\w^* )$, 
\begin{small}
\begin{equation}
\begin{split}
    \Psi = \frac{1}{c}&\sum_{m\in\mathcal{M}} \bigg[\frac{L\eta_g \eta_l}{2MN_m}\sigma_{m}^2 \\
    & + \sum_{\theta\subseteq 2^\mathcal{M}}\frac{5N_{m,\theta}EL^2 \eta_l^2}{2MN_m}\bigg(\sigma_m^2 + \frac{6EN_{m,\theta}}{MN_m}\alpha_{m,\theta}^2 \bigg)\bigg].\nonumber
\end{split}
\end{equation}
\end{small}
\end{theorem}
\noindent\textit{Proof.} The proof is shown in Appendix~B.\hfill~$\Box$

\textbf{Remark 1.} For the full client participation strategy of MS-FedAvg algorithm, the convergence rate has two parts: a vanishing term $\frac{f_0 - f_*}{cMET\eta_g \eta_l}$ with increasing $T$ and a constant term $\Psi$. The first part of $\Psi$, i.e., $\frac{L\eta_g \eta_l}{2MN_m}\sigma_{m}^2$, comes from the local stochastic gradient variance of each client, which shrinks when $N_m$ increases. The cumulative variance of $E$ local training contributes to the second term of $\Psi$, which has two variances and is largely affected by variance of different regional models $\alpha_{m,\theta}^2$. 

\textbf{Remark 2.} The difference between MS-FedAvg and single server FedAvg \cite{reddi2020adaptive,yang2021achieving} comes from the term $\alpha_{m,\theta}^2$. Since the initial learning models $\w_i^t$ are the same for all clients, $\alpha_{m,\theta}^2$ is only related to the non-iid distribution of local datasets, i.e., $\alpha_{m,\theta}^2 = \alpha_m^2$, and the weight is all the same $\frac{1}{N_m}$. In MS-FedAvg, we observe that the contribution of $\alpha_{m,\theta}^2$ depends on the number of clients in each area type $\theta$, i.e., $\frac{6EN_{m,\theta}^2}{N_m^2}\alpha_{m,\theta}^2$. Intuitively, local model from clients of the area type with the most clients can dominate the regional model $\w_m^t$. Inspired by \cite{yang2021achieving}, to make $\Psi$ small, we can set the local learning rate $\eta_l$ inversely proportional to the number of local training epochs $E$, i.e., $\eta_l = O(\frac{1}{E})$.

To make the Theorem~\ref{The:nonconvexfull} more readable, we will simplify the result to the following convergence rate by properly choosing the learning rates $\eta_g$ and $\eta_l$:

\begin{corollary}\label{Cor:full}
Suppose $\eta_g$ and $\eta_l$ satisfy the condition in Theorem~\ref{The:nonconvexfull}. Let $\eta_g = \sqrt{EN_m}$ and $\eta_l = \frac{1}{\sqrt{T}EL}$. For sufficiently large $T$, the convergence rate of MS-FedAvg under full client participation strategy satisfies:
\begin{small}
\begin{equation}
\begin{split}
    & \min_{t \in [T]}\mathbb{E}||\nabla f (\w^t )||^2 = \\
    &O\bigg(\frac{1}{M}\sum_{m\in\mathcal{M}}\bigg(\frac{1}{\sqrt{N_m ET}}+ \frac{\sigma_m^2}{ET} +\sum_{\theta \subseteq 2^\M}\frac{N_{m,\theta}^2 \alpha_{m,\theta}^2}{N_m^2}\frac{1}{T}\bigg)\bigg).\nonumber
\end{split}
\end{equation}
\end{small}
\end{corollary}

\subsection{Convergence Analysis for Unbiased Partial Client Participation}
Due to the limited resource for current FL wireless networks, partial participation strategy (only part of clients join into the current communication round) has been considered more practical than full participation in existing FL studies \cite{li2019convergence, karimireddy2020scaffold, qu2020federated}. Also, partial participation can accelerate the training by neglecting stragglers. We also consider the same two sampling schemes \cite{li2019convergence,qu2020federated,yang2021achieving} for MS-FedAvg algorithm, i.e., with/without replacement clients sampling schemes, where $\mathcal{K}_m^t$ is randomly sampled. Due to the random property, we call it \textbf{unbiased client participation} of our proposed multi-server FL in this paper. More specifically, it is worth noting that the unbiased client participation strategy for MS-FedAvg implies that $\frac{\E[K_{m, \theta}]}{K_m}=\frac{N_{m,\theta}}{N_m}$. Then, we have the following convergence results:

\begin{theorem}\label{The:partial}
Let assumptions~\ref{Ass:Lsmooth}-\ref{Ass:regionalbound} hold and $L, \sigma_{m}^2, \alpha_{m,\theta}^2$ be defined therein. Let $\beta_{m,\theta}^2 = \sigma_m^2 + \frac{6EN_{m,\theta} \alpha_{m,\theta}^2}{N_m}$. With the scheme I for the unbiased client participation strategy, if the learning rate is chosen as $\eta_l < \min\left\{\frac{1}{\sqrt{30}EL}, \sum_{\theta \subseteq 2^\M}\frac{N_{m,\theta}(K_{m,\theta}-1)}{EL \eta_g K_m N_m}\right\}$ and the condition $30E^2L^2\eta_l^2+\frac{L\eta_g \eta_l \sum_{\theta \subseteq 2^\M} N_{m,\theta}}{K_m N_m}(90E^3 L^2 \eta_l^2 +3E) < 1$ holds, the global model $\w^T$ generated by MS-FedAvg in Algorithm~\ref{ALG:MSFL} satisfies:
\begin{small}
\begin{equation}
    \min_{t \in [T]} \mathbb{E}||\nabla f (\w^t )||^2 \leq \frac{f^0 - f^*}{cM\eta_g \eta_l ET} + \Psi_1 + \Psi_2 +\Psi_3 \nonumber,
\end{equation}
\end{small}
where $c$ is a constant, $f^0 \triangleq f (\w^0 )$ and $f^* \triangleq f (\w^* )$, 
\begin{small}
\begin{equation}
\begin{split}
    &\Psi_1 = \sum_{m\in\mathcal{M}}\frac{EL\eta_g \eta_l}{2cMK_m}\sigma_m^2 , \Psi_2 =  \sum_{\theta \subseteq 2^\M}\frac{3EL\eta_g \eta_l N_{m,\theta}}{2cMK_m N_m}\alpha_{m,\theta}^2 \\
    &\Psi_3 =\sum_{m\in\mathcal{M}}\sum_{\theta \subseteq 2^\M}\bigg(\frac{5N_{m,\theta}EL^2 \eta_l^2}{2cMN_m} +\frac{15N_{m,\theta}E^2 L^3 \eta_g\eta_l^3}{cMK_m N_m}\bigg)\beta_{m,\theta}^2 .\nonumber
\end{split}
\end{equation}
\end{small}

For the Scheme II, if the learning rate is chosen as $\eta_l <\min\left\{ \frac{1}{\sqrt{30}EL}, \sum_{\theta \subseteq 2^\M}\frac{K_{m}^2N_{m,\theta}(N_{m,\theta}-1)}{EL \eta_g N_m^2 K_{m,\theta}(K_{m,\theta}-1)}\right\}$ and the condition $30E^2L^2\eta_l^2 +\sum_{\theta \subseteq 2^\M} \frac{L\eta_g \eta_l(K_{m,\theta}-1)}{2K_m (N_{m,\theta}-1)}(90E^3 L^2\eta_l^2 + 3E) < 1$ holds, then we obtain that 
\begin{small}
\begin{equation}
\begin{split}
    \Psi_1 = \sum_{m\in\mathcal{M}}&\frac{L\eta_g \eta_l}{2cMK_m}\sigma_m^2 \\
    \Psi_2 =\sum_{m\in\mathcal{M}}&\sum_{\theta \subseteq 2^\M}\frac{2EL^2 \eta_g\eta_l N_{m,\theta}(N_{m,\theta}-K_{m,\theta})}{cMK_m N_m (N_{m,\theta}-1)}\alpha_{m,\theta}^2 \\
    \Psi_3 = \sum_{m\in\mathcal{M}}& \sum_{\theta \subseteq 2^\M}\bigg(\frac{5EL^3 \eta_l^2 N_{m,\theta}}{2cMN_m}\\
    &+\frac{15E^2 L^3 \eta_g \eta_l^3 N_{m,\theta}(N_{m,\theta}-K_{m,\theta})}{2cMN_m K_m (N_{m,\theta}-1)}\bigg)\beta_{m,\theta}^2 .\nonumber
\end{split}
\end{equation}
\end{small}
\end{theorem}

\noindent\textit{Proof.} The proof is shown in Appendix~C. \hfill~$\Box$

Similar to full participation, we restate the above result by properly choosing $\eta_g$ and $\eta_l$:
\begin{corollary}\label{cor:partial}
Suppose $\eta_g$ and $\eta_l$ satisfy the condition in Theorem~\ref{The:partial}. Let $\eta_g = \sqrt{EK_m}$ and $\eta_l = \frac{1}{\sqrt{T}EL}$. Then, for sufficiently large $T$, the convergence rate of MS-FedAvg under unbiased partial client participation strategy satisfies:
\begin{small}
\begin{equation}
\begin{split}
    &\min_{t \in [T]}\mathbb{E}||\nabla f_m (\w_t )||^2 = O\bigg(\frac{1}{M}\sum_{m\in\mathcal{M}}\bigg( \frac{1}{\sqrt{K_m E T}} + \frac{\sigma_m^2}{ET} \\
    &+ \sum_{\theta \subseteq 2^\M} \frac{N_{m,\theta}^2 \alpha_{m,\theta}^2}{N_m^2}\frac{\sqrt{E}}{\sqrt{K_m T}} + \sum_{\theta \subseteq 2^\M}\frac{N_{m,\theta}^2 \alpha_{m,\theta}^2}{N_m^2}\frac{1}{T}\bigg) \bigg).\nonumber
\end{split}
\end{equation}
\end{small}
\end{corollary}

\textbf{Remark 3.} The structure of the convergence rate of MS-FedAvg algorithm under unbiased partial client participation strategy is similar to full client participation, except an additional variance term $\Psi_2$. This indicates that the unbiased partial client participation strategy does not have significant change in convergence except for an amplified variance due to fewer clients being sampled. Intuitively, it yields a good approximation of all the clients' datasets distribution in expectation.

\textbf{Remark 4.} From Corollary~\ref{cor:partial}, we can see that the convergence rate of the unbiased partial participation strategy is not related to the number of clients in each area type $K_{m,\theta}$, but it highly depends on the ratio $\frac{N_{m,\theta}^2}{N_m^2}$. However, due to the complicated network topology of multi-server FL, clients in some area types may present extreme performance, e.g., large $\alpha_{m,\theta}^2$ to incur large training degradation. Hence, to further accelerate the convergence rate of MS-FedAvg, we will develop a new sampling strategy that samples different numbers of clients in different area types. 

\begin{table*}[t!]
\centering
\caption{Convergence rate of existing benchmarks.}
\adjustbox{width=1.85\columnwidth}{\begin{threeparttable}
\begin{tabular}{cccccc}
\hline
Algorithm & Network architecture             & Convexity\tnote{1} &  Assumptions\tnote{2}     & Partial client       & Convergence rate                        \\ \hline
FedAvg \cite{li2019convergence}   & Single server                 & SC    &  BGD  & $\checkmark$ & $O(\frac{E}{T})$                        \\
FedAvg \cite{yang2021achieving}   & Single server                 & NC    &  BGV  & $\checkmark$ & $O(\frac{1}{\sqrt{NET}} + \frac{1}{T})$ \\
MC-PSGD \cite{ding2020distributed}  & Cluster                       & SC   & BGD; BMP       & $\times$     & $O(\frac{1}{\sqrt{N_m T}})$                \\
IFCA \cite{ghosh2020efficient}     & Cluster                       & SC   &  BGD   & $\checkmark$ & $O(\frac{1}{\sqrt{N_m T}} + \frac{E}{T})$  \\
HFL     \cite{wang2020local}  & Hierarchical                  & NC     & BGV; BLV  & $\times$     & $O(\frac{1}{\sqrt{N_m T}}+\frac{1}{T})$    \\
FedMes  \cite{han2021fedmes} & Multi-server with overlapping areas & SC     &  BGD & $\times$ &   $O(\frac{KE^2}{N})$                                \\
MS-FedAvg & Multi-server with overlapping areas & NC    &  BGV  & $\checkmark$ & $O(\frac{1}{\sqrt{N_m ET}}+\frac{1}{T})$  \\ \hline
\end{tabular}
\begin{tablenotes}\footnotesize
\item[1] Shorthand notation for convexity: SC: Strongly Convex and NC: Non-Convex.
\item[2] Shorthand notation for assumptions in the paper: BGD is bounded gradient descent $\|\nabla f(\w) \|^2 \leq G^2$; BGV is bounded global variance $\|\nabla f_i (\w) - \nabla f(\w)\| \leq \sigma^2$; BMP is bounded model parameter $\|\w\|^2 \leq B^2$; BLV is bounded local variance $\|\nabla f_i (\w) - \nabla f_j (\w)\|^2 \leq \epsilon^2$.
\end{tablenotes}
\end{threeparttable}}
\label{Tab:convergence}
\end{table*}

\subsection{Convergence Analysis of Biased Partial Client Participation}
In this subsection, we aim to develop a new \textbf{biased partial client participation} to achieve speedup of the MS-FedAvg algorithm. Let $\K_{m,\text{biased}}^t \subseteq \N_m$ be the sampled clients set based on this strategy with $K_m =|\K_{m,\text{biased}}^t |$. The main idea of this strategy is that the number of sampled clients $\E [K_{m,\theta}]$ in different area type $\theta$ is fixed, where the ratio $\frac{\E[K_{m,\theta}]}{K_m}$ may not be equal to $\frac{N_{m,\theta}}{N_m}$.  $\frac{\E[K_{m,\theta}]}{K_m}$ reflects the degree of bias. Intuitively, we can reduce the sampling number $\E[K_{m,\theta}]$ for some area types with large $\alpha_{m,\theta}^2$ in order to reduce their convergence contribution. Note that this strategy also includes the same two schemes with/without replacement as the unbiased participation strategy. The convergence results are shown as follows:
\begin{theorem}\label{The:biasedpartial}
Let assumptions~\ref{Ass:Lsmooth}-\ref{Ass:regionalbound} hold and $L, \sigma_{m}^2, \alpha_{m,\theta}^2$ be defined therein, and $c$ is a constant. Let $\beta_{m,\theta}^2 = \sigma_m^2 + \frac{6EK_{m,\theta}}{K_m}$. With scheme I for the biased client participation strategy, if the learning rate is chosen as $\eta < \min\left\{\frac{1}{\sqrt{30}ET}, \sum_{\theta \subseteq 2^{\mathcal{M}}}\frac{K_{m,\theta}(K_{m,\theta}-1)}{EL\eta_g K_m^2 }\right\}$ and the condition $30E^2L^2 \eta_l^2 +\sum_{\theta \subseteq 2^\M}\frac{L\eta_g \eta_l K_{m,\theta}^2}{K_m^2 N_{m,\theta}}(90E^3 L^2 \eta_l^2 + 3E) < 1$ holds, the global model $\w^T$ generated by MS-FedAvg in Algorithm~\ref{ALG:MSFL} satisfies:
\begin{small}
\begin{equation}
    \min_{t \in [T]} \mathbb{E}||\nabla f (\w^t )||^2 \leq \frac{f^0 - f^*}{cM\eta_g \eta_l ET} + \Psi_1 + \Psi_2 + \Psi_3 \nonumber,
\end{equation}
\end{small}
where $c$ is a constant, $f^0 \triangleq f(\w^0 )$ and $f^* \triangleq f(\w^* )$, 
\begin{small}
\begin{equation}
\begin{split}
    \Psi_1 = \sum_{m\in\mathcal{M}}& \frac{L\eta_g\eta_l}{2cMK_m}\sigma_m^2 \\
    \Psi_2 = \sum_{m\in\mathcal{M}}\sum_{\theta \subseteq 2^\M}&\frac{3EL\eta_g\eta_l K_{m,\theta}^3}{2cMK_m^3 N_{m,\theta}}\alpha_{m,\theta}^2 \\
    \Psi_3 = \sum_{m\in\mathcal{M}}\sum_{\theta \subseteq 2^\M}&\bigg(\frac{5EL^2 \eta_l^2 K_{m,\theta}}{2cMK_{m}} +\frac{15E^2L^3 \eta_g\eta_l^3 K_{m,\theta}^2 }{2cMK_m^2 N_{m,\theta}}\bigg)\beta_{m,\theta}^2 .\nonumber
\end{split}
\end{equation}
\end{small}

For Scheme II, if the learning rates is chosen as $\eta_l < \min\left\{ \frac{1}{\sqrt{30}KL}, \sum_{\theta \subseteq 2^\M}\frac{K_{m,\theta}^2 (N_{m,\theta}-1)}{EL\eta_g N_{m,\theta} K_{m}(K_{m,\theta}-1)}\right\}$ and the condition $30E^2 L^2\eta_l^2+ \sum_{\theta \subseteq 2^\M}\frac{L\eta_g \eta_l K_{m,\theta} (N_{m,\theta} - K_{m,\theta})}{K_m N_{m,\theta}(N_{m,\theta} - 1)}(90E^3 L^2 \eta_l^2 +3E) < 1$ holds, and then we obtain that 
\begin{small}
\begin{equation}
\begin{split}
    \Psi_1 & = \sum_{m\in\mathcal{M}}\frac{L\eta_g \eta_l}{2cMK_m}\sigma_m^2 \\
    \Psi_2 & = \sum_{m\in\mathcal{M}}\sum_{\theta \subseteq 2^\M}\frac{3L\eta_g \eta_l K_{m,\theta}(N_{m,\theta}-K_{m, \theta})}{2cK_m^2 N_{m,\theta}(N_{m,\theta}-1)}\alpha_{m,\theta}^2 \nonumber
\end{split}
\end{equation}
\begin{equation}
\begin{split}
    \Psi_3 = EL^2 &\eta_l^2 \sum_{m\in\mathcal{M}}\sum_{\theta \subseteq 2^\M}\bigg(\frac{5K_{m,\theta}EL^2 \eta_l^2}{2cMK_m}\\
    & + \frac{15E^2 L^3 \eta_g \eta_l^3 K_{m,\theta} (N_{m,\theta}-K_{m,\theta})}{2cMK_m N_{m,\theta}(N_{m,\theta}-1)}\bigg)\beta_{m,\theta}^2 .\nonumber
\end{split}
\end{equation}
\end{small}
\end{theorem}
\noindent\textit{Proof.} The proof is shown in Appendix~D.\hfill~$\Box$

\begin{corollary}\label{cor:biased}
Suppose $\eta_g$ and $\eta_l$ satisfy the condition in Theorem~\ref{The:partial}. Let $\eta_g = \sqrt{EK_m}$ and $\eta_l = \frac{1}{\sqrt{T}EL}$. Then, for sufficiently large $T$, the convergence rate of MS-FedAvg under biased partial client participation strategy satisfies:
\begin{small}
\begin{equation}
\begin{split}
    &\min_{t \in [T]}\mathbb{E}||\nabla f_m (\w_t )||^2 = O\bigg(\frac{1}{M} \sum_{m\in\mathcal{M}}\bigg(\frac{1}{\sqrt{K_m E T}} + \frac{\sigma_m^2}{ET}\\
    &+ \sum_{\theta \subseteq 2^\M}\frac{K_{m,\theta}^2 \alpha_{m,\theta}^2}{K^2_m}\frac{\sqrt{E}}{\sqrt{K_m T}} + \sum_{\theta \subseteq 2^\M}\frac{K_{m,\theta}^2 \alpha_{m,\theta}^2}{K_m^2}\frac{1}{T}\bigg) \bigg).\nonumber   
\end{split}
\end{equation}
\end{small}
\end{corollary}

\textbf{Remark 5.} From Corollary~\ref{cor:biased}, we can see that the biased client participation strategy has the same structure as the unbiased strategy. The difference is that variances of $\alpha_{m,\theta}^2$ include the term $\frac{K_{m,\theta}^2}{K_m^2}$ not $\frac{N_{m,\theta}^2}{N_m^2}$. Obviously, it is not difficult to design $K_{m,\theta}$ for each area type to accelerate convergence, for example, we can sample more clients in some areas with lower $\alpha_{m,\theta}^2$ value (suppose that $\alpha_{m,\theta}^2$ is constant) in order to decrease the variance terms $\Psi_2$ and $\Psi_3$. More specifically, since the variance $\alpha_{m,\theta}^2$ should be related to $K_{m,\theta}$, i.e., increasing $K_{m,\theta}$ should decrease $\alpha_{m,\theta}^2$, if the sampling strategy is such a way, it achieves a significant speedup for convergence in training.

\section{Discussion of MS-FedAvg}\label{Sec:Discussion}
Based on the above results, we briefly discuss the theoretical analysis of MS-FedAvg and its implications.

\textbf{Convergence Rate:} When $T$ is sufficiently large compared to $E$, we can simplify the convergence rates $O(\frac{1}{\sqrt{N_m ET}} + \frac{1}{T})$ in Corollary~\ref{Cor:full} and $O(\frac{\sqrt{E}}{\sqrt{K_m T}}+ \frac{1}{T})$ in Corollaries~\ref{cor:partial}-\ref{cor:biased}, which matches the rate in the general non-convex setting of single-server FL algorithms \cite{karimireddy2020scaffold, reddi2020adaptive, yang2021achieving} without the consideration of transmission difference. Although some works proposed new algorithms for multi-server FL architectures \cite{duan2020self, lee2020tornadoaggregate, ghosh2020efficient, xie2020multi}, few of them presented the detailed convergence analysis. In Table~\ref{Tab:convergence}, we summarize the convergence rate of some existing FL studies. Compared to the convergence rate, it is easy to see that our proposed MS-FedAvg algorithm achieves linear speedup for general non-convex settings. More specifically, our assumption is the most strict among these studies, and BMP assumption should be unrealistic.

\textbf{Accuracy:} Although theorem~\ref{The:biasedpartial} shows that sampling clients from fewer area types can improve the convergence performance, if we miss the clients in some area types, the accuracy performance may be degraded due to overfitting. Therefore, the design condition is that $\E[K_{m,\theta}] >0$, $\forall \theta$. Due to the complicated network topology of multi-server FL architecture, it is difficult to obtain the theoretical result of $K_{m,\theta}$. We will present the empirical results to support our accuracy discussion of the biased partial client participation strategy in the Section~\ref{Sec:experiment}.

\textbf{Number of Local epochs $E$ and Client $K_m$:} Our results show that the number of local training epochs can be set as $E \leq \frac{T}{K_m}$ to accelerate convergence. We also show that the local training epochs help the convergence by properly setting hyper-parameters, which supports the previous results \cite{mcmahan2017communication,stich2018local,yang2021achieving}. The results in Theorems~\ref{The:nonconvexfull}-\ref{The:biasedpartial} imply that the convergence rate can be improved substantially by increasing the number of clients in each communication round.

{\hlb \textbf{Comparisons to FedMes \cite{han2021fedmes}:} Although the training procedure of MS-FedAvg is similar to FedMes in \cite{han2021fedmes}, the unique difference between these two algorithms is that our proposed MS-FedAvg can leverage the value of $\eta_g$, which has been demonstrated that finding an optimal $\eta_g$ can accelerate the training performance \cite{reddi2020adaptive, yang2021achieving}. In Table~\ref{Tab:convergence}, we can see that \cite{han2021fedmes} only proposes the convex loss function of FedMes (e.g., logistic regression \cite{li2019convergence}). Although \cite{han2021fedmes} presented the experimental results based on CNN model and achieve improvement, it does not propose the theoretical analysis to support the result. Since most of existing machine learning algorithms are non-convex (e.g., CNN and LSTM \cite{reddi2020adaptive, yang2021achieving, karimireddy2020scaffold}), the theoretical results in \cite{han2021fedmes} is much more restricted. In this paper, the theoretical analysis and experiments are both on the general non-convex settings. In addition, the convergence analysis in \cite{han2021fedmes} leverages the BGD assumption, which has been considered a loose assumption in existing FL studies \cite{yang2021achieving}. As such, our convergence analysis is tighter than FedMes. Lastly, we propose two kinds of partial client participation strategies (each strategy has two sampling schemes), and analyze the training performance based on the ratio of the number of clients in different area types, which did not mention in \cite{han2021fedmes}.

\textbf{Limitations:} The regional models in MS-FedAvg are not aggregated before the $T$th round. Hence, the final round aggregation does not have significant impact on the convergence. The implicit aggregation is due to the fact that the clients in overlapping areas share information across all regional models. Considering all factors in MS-FedAvg, including architecture, client distribution and heterogeneous local dataset makes the contribution of the implicit aggregation to be captured difficultly so that the full analysis mathematically intractable. Thus, we bound the factors that depend on the convergence results between different servers via Assumption~\ref{Ass:regionalbound}, and analyze the convergence in each region. As such, the problem becomes tractable and at the same time does not substantially impact the final results. In the future, we will set the multi-server FL as a bipartite graph, and propose the consensus analysis (i.e., the convergence gap between regional models and global model).}

\section{Transmission Latency Analysis}\label{Sec:latency}
{\hlb \textbf{1) MS-FedAvg:} In the multi-server FL network, to calculate the running time $\tau_{\text{Multi}} (t)$ in every communication round $t$, we will present the expressions to compute the three main components local computing time $\tau_i^{C} (t)$, uploading time $\tau_{i,m}^{U}(t)$ and downloading time $\tau_{i,m}^{D}(t)$. Note that because our proposed algorithms mainly focus on the efficiency of transmission, and the local computing time $\tau_i^C (t)$ is negligible compared to transmission latency \cite{konevcny2016federated, kairouz2019advances}, we omit this part in our experiments. In summary, the transmission latency $\tau_{\text{Multi}}(t)$ in each round $t$ is the sum of the largest uploading time and downloading time, i.e.,
\begin{small}
\begin{equation}\label{Eq:TimeMulti}
    \tau_{\text{Multi}} (t) = \max_{i,m}\tau_{i,m}^U (t) + \max_{i,m}\tau_{i,m}^D (t). 
\end{equation}
\end{small}
Uploading time of client $i$ in communication round $t$ is defined as follows:
\begin{small}
\begin{equation}
    \tau_{i,m}^U (t) = \frac{q_i}{b r_{i,m}^U (t)},
\end{equation}
\end{small}
where $q_i$ is the data size of client $i$ for uploading and $r_{i,m}^U (t)$ in bits/s/Hz denotes the uploading rate of client $i$ to the corresponding regional server $m$ in communication round $t$, which is defined as follows:
\begin{small}
\begin{equation}
    r_{i,m}^U (t) = \log_2 \bigg(1+\frac{p_{i,m}^U |g_{i,m}^U (t)|^2}{\mu^2}\bigg),
\end{equation}
\end{small}
where $p_{i,m}^U$ is the uplink transmit power of and $g_{i,m}^U (t)$ is the uplink channel gain of client $i$ to the corresponding regional server $m$ in communication round $t$, and $\mu^2$ is the channel noise. Note that $b$ in Hz is the bandwidth of one channel, i.e., $b = B/N$, where $B$ is the total bandwidth budget and $N$ is the number of clients. If we use partial participation strategy $b = B/N$. Since our compared benchmarks include multiple different FL network architectures, bandwidth $b$ is divided into three categories: (1) $b_{cr} = B_{cr} / N$ is the client to regional server bandwidth; (2) $b_{rc} = B_{rc} /N$ is the regional server to cloud server bandwidth and (3) $b_{cc} = B_{cc} / N$ is the client to cloud server bandwidth. In the real world mobile network, $b_{cr} \leq b_{rc} = b_{cc}$ \cite{mao2017survey}.

The definition of downloading time of client $i$ to the corresponding regional server $m$ is $\tau_{i,m}^D (t)$ is similar to the uploading time $\tau_{i,m}^U (t)$, which is defined as $\tau_{i,m}^D (t) = \frac{q_{i,m}}{b r_{i,m}^D (t)}$, where $r_{i,m}^D (t) = \log_2 (1+\frac{p_{i,m}^D |g_{i,m}^D (t)|^2}{\mu^2})$, $p_{i,m}^D$ is the downlink transmit power, $g_{i,m}^D (t)$ is the downlink channel gain of client $i$ to the corresponding regional server $m$ in communication $t$. Suppose that the total communication round to achieve the targeted testing accuracy is $T_{\text{Multi}}$, the total transmission time is $\tau^{\text{Total}}_{\text{Multi}} = \sum_{t=1}^{T_{\text{Multi}}}\tau (t)$. Specifically, the transmission latency calculation of FedMes \cite{han2021fedmes} is the same as MS-FedAvg.}

{\hlb \textbf{2) Single-server FL:} In the single-server FL network architecture, all clients communicate to the central server for download/uploading the model updates uploading/downloading. The transmission latency $\tau_{\text{Single}}(t)$ of the single-server FL for one communication round depends on the slowest client $i$, which is calculated by
\begin{small}
\begin{equation}
    \tau_{\text{Single}} (t) = \max_{i}\tau_{i}^U + \max_{i}\tau_{i}^D .
\end{equation}    
\end{small}
Note that the transmit power $p_i^D$ and $p_i^U$ and the channel gain $g_i^D$ and $g_i^U$ should decay with increasing the distance \cite{liu2019dynamic, tran2018joint}. Clearly, the distance between clients and regional server(s) should be much less than the distance between clients and central server. Even though many existing single-server FL studies have proposed the developed algorithm to improve the convergence rate \cite{li2019convergence, karimireddy2020scaffold, karimireddy2021breaking}, single-server FL also requires much more total transmission time due to the large value of $\tau_{\text{Single}} (t)$. The total transmission time of single-server FL is $\tau^{\text{Total}}_{\text{Single}} = \sum_{t=1}^{T_{\text{Single}}}\tau (t)$.

\textbf{3) HFL:} The HFL architecture includes both the regional servers and the central server \cite{lim2020federated, liu2020client, wang2020local}, which has two aggregation schemes (i.e., edge aggregation and global aggregation). In each region aggregation round, each regional server aggregates the local model updates uploading from the clients in its service area, where the transmission latency of region aggregation is 
\begin{small}
\begin{equation}
    \tau_{\text{Region}} (t) = \max_{i,m}\tau_{i,m}^U (t) + \max_{i,m}\tau_{i,m}^D (t) . 
\end{equation}
\end{small}
In the global aggregation round, the central server aggregates the model updates on each regional server in which the transmission latency is
\begin{small}
\begin{equation}
    \tau_{\text{Global}} (t) = \max_{m}\tau_{m}^U (t) + \max_{m}\tau_{i,m}^D (t) . 
\end{equation}
\end{small}
Note that the global aggregation round is performed periodically at every $t_{\text{Global}}$ edge aggregation round (i.e., $t_{\text{Global}} \geq 1$). Suppose that if HFL requires $T_{\text{Region}}$ and $T_{\text{Global}}$ to achieve the targeted accuracy, the total transmission time of HFL is $\tau^{\text{Total}}_{\text{HFL}} = \sum_{t=1}^{T_{\text{Region}}}\tau_{\text{Region}}(t) + \sum_{t' =1}^{T_{\text{Global}}}\tau_{\text{Global}}(t')$. Clearly, we can observe that HFL has extra aggregation rounds (i.e., global aggregation round) compared to single-server FL and multi-server FL architectures from which $\tau_{\text{Global}} > \tau_{\text{Region}}$ due to the large distance between regional servers and central server. More specifically, in Table~\ref{Tab:convergence}, the convergence rate of HFL (i.e., $\frac{1}{\sqrt{N_m T}} + \frac{1}{T}$), which implies that HFL requires more communication round to achieve targeted accuracy and incurs large total transmission time.

\textbf{4) CFL:} We assume that the CFL architecture includes $M$ regional servers, where the number of $M$ is equal to the number of clusters. Although the calculation of one communication round transmission latency of CFL $\tau_{\text{CFL}}$ is the same as the multi-server FL in \eqref{Eq:TimeMulti}, the distance of clients to regional server is usually larger than multi-server FL since the clustering policy aims to cluster the clients that perform similar dataset distribution \cite{ghosh2020efficient, duan2020fedgroup}. In addition, the capability of regional server is much lower than central server, and hence the slowest client will high impact on the transmission latency (i.e., $\tau_{\text{CFL}} \gg \tau_{\text{Multi}}$). Specifically, CFL should re-cluster the clients after every several communication rounds, which incurs extra communication latency. The CFL may incur high divergence of each cluster, which may degrade the convergence performance, and it should use more communication rounds to achieve the targeted accuracy. In summary, if the number of total communication round is $T_{\text{CFL}}$ and the number of re-clustering is $T_{\text{Cluster}}$, the total transmission time of CFL is $\tau_{\text{Total}}^{\text{CFL}} = \sum_{t=1}^{T_{\text{CFL}}}\tau_{\text{CFL}}(t) + \sum_{t'=1}^{T_{\text{Cluster}}}\tau_{\text{Cluster}}(t')$.}

\begin{table*}[t!]
\centering
\caption{Datasets and models.}
\adjustbox{width=1.85\columnwidth}{\begin{tabular}{cccccc}
\hline
Dataset & Task & Clients & Total samples & Batch size & Model \\ \hline
EMNIST \cite{cohen2017emnist}                        & Handwritten character recognition &  85     & 81,425 & 16 & 2-layer CNN+2-layer FFN\\
CIFAR-10 \cite{krizhevsky2009learning} & Image classification &  85 & 60,000 & 32 & MobileNet-v2 \cite{sandler2018mobilenetv2} \\ 
CIFAR-100 \cite{krizhevsky2009learning} & Image classification &  85 & 60,000 &32 & MobileNet-v2 \cite{sandler2018mobilenetv2} \\\hline
\end{tabular}}
\label{Tab:model}
\end{table*}

\section{Experiments}\label{Sec:experiment}
\subsection{Experimental Setup}
\textbf{Datasets and models.} We evaluate our proposed algorithms on three datasets: EMNIST \cite{cohen2017emnist}, CIFAR-10 and CIFAR-100 \cite{krizhevsky2009learning}. In each dataset, we simulate the data heterogeneity by sampling the label ratios from a Dirichlet distribution with parameter 0.4 \cite{hsu2019measuring}, and keep the training data on each client balanced. For EMNIST dataset, we use CNN model with two hidden-layers and two FeedForward Network (FFN) layers, and the two learning rates are set as $\eta_g =1.1$ and $\eta_l = 0.05$ by grid search. For CIFAR-10 and CIFAR-100, we use MobileNet-v2 \cite{sandler2018mobilenetv2} to be the learning model, and the learning rates are set as $\eta_g = 1.5$ and $\eta_l = 0.1$. Table~\ref{Tab:model} summarizes datasets, models, batch sizes and the number of clients. {\hlb All the hyper-parameters are set based on grid search on each dataset.} Note that all the algorithms are set $E=5$ and $K_m =10$ by default.

\begin{figure}[t!]
    \centering
     \includegraphics[width=\FigWidthSmaller]{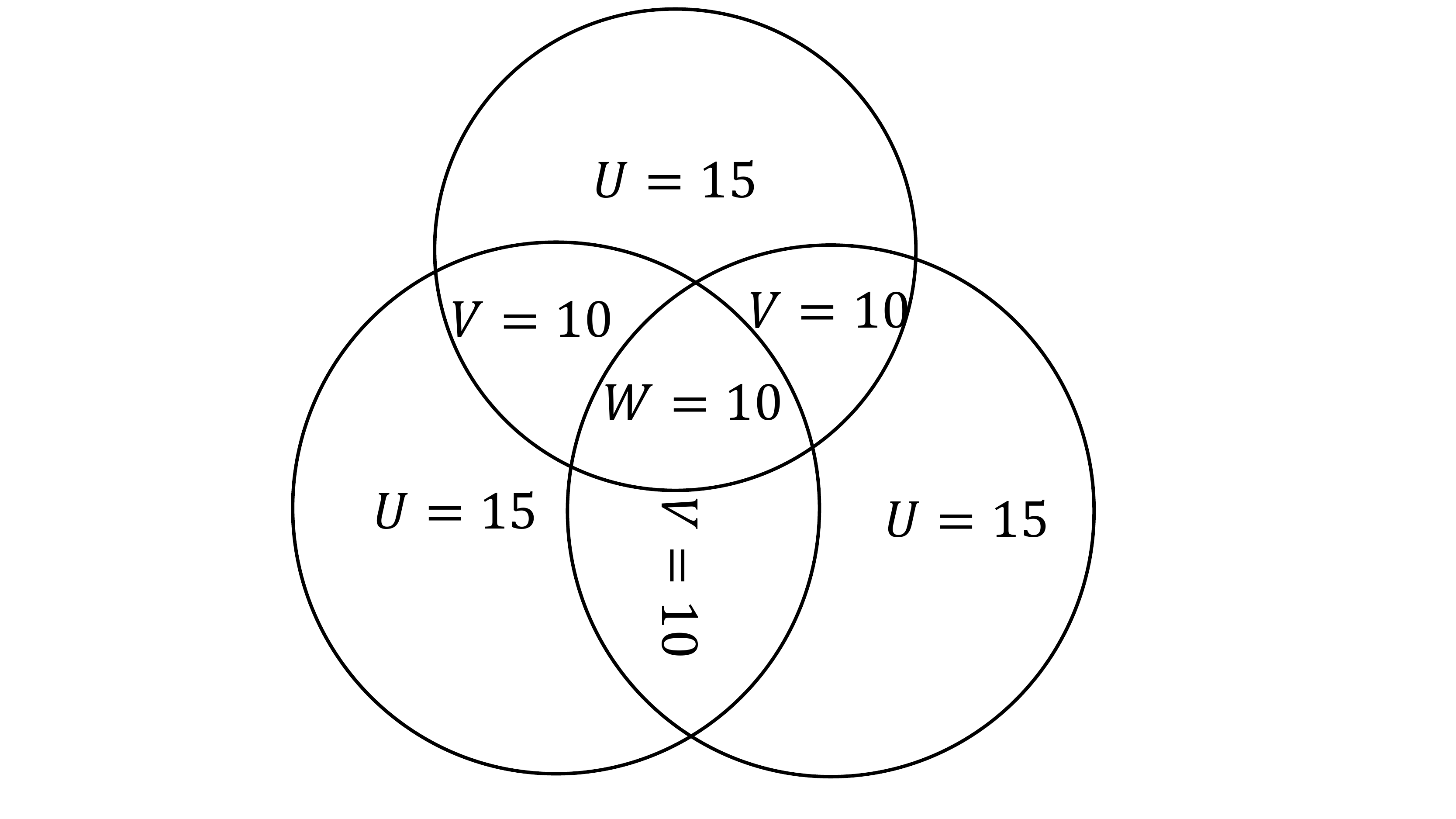}
    \caption{Symmetric multi-server FL architecture.}
    \label{Fig:symmetric}
\end{figure}

\begin{figure*}[t!]
\centering
\begin{minipage}{0.64\columnwidth}
  \centering
\includegraphics[width=\textwidth, height=0.7\columnwidth]{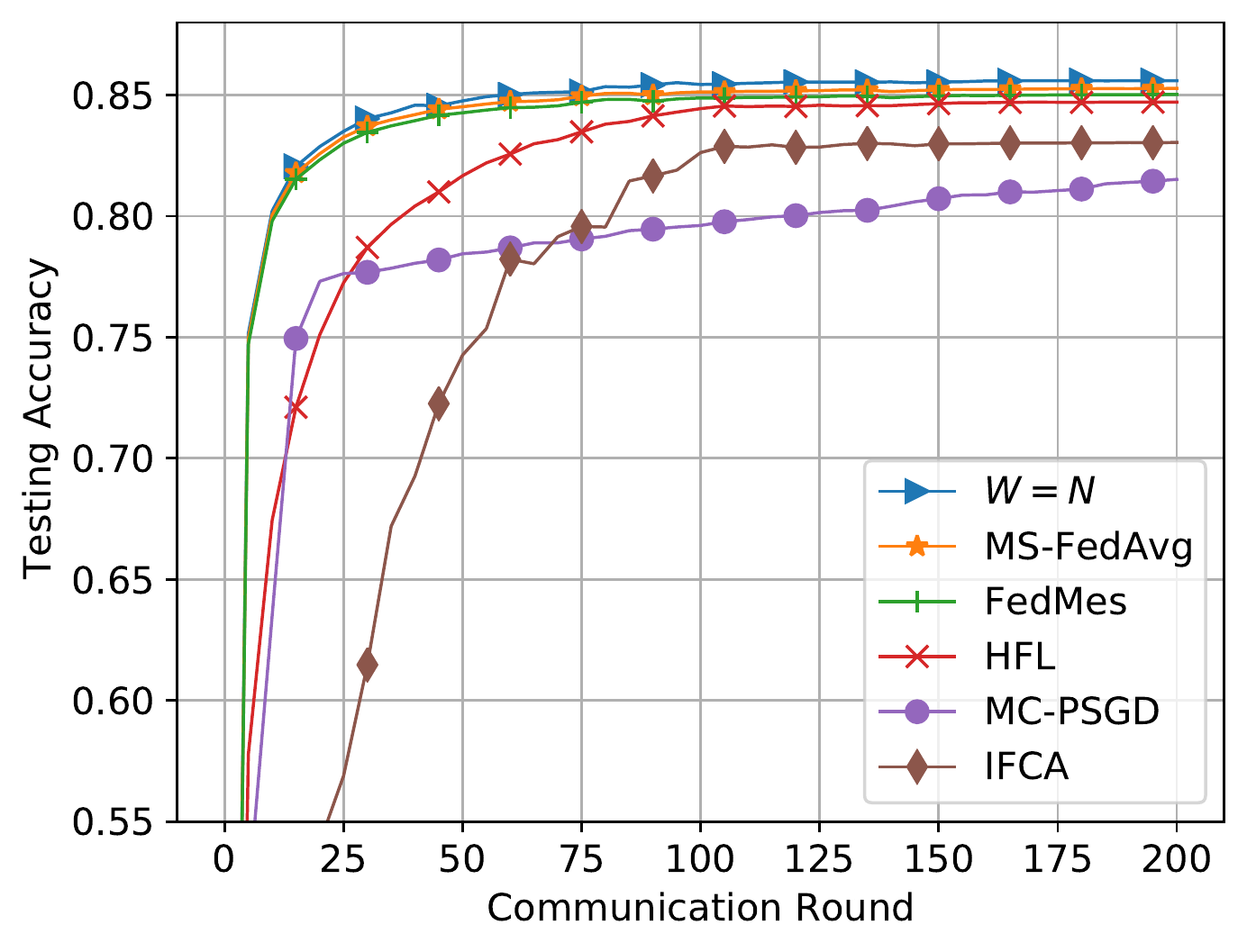}
\subcaption[first]{EMNIST dataset.}
\end{minipage}
\hfill
\begin{minipage}{0.64\columnwidth}
  \centering
\includegraphics[width=\textwidth, height=0.7\columnwidth]{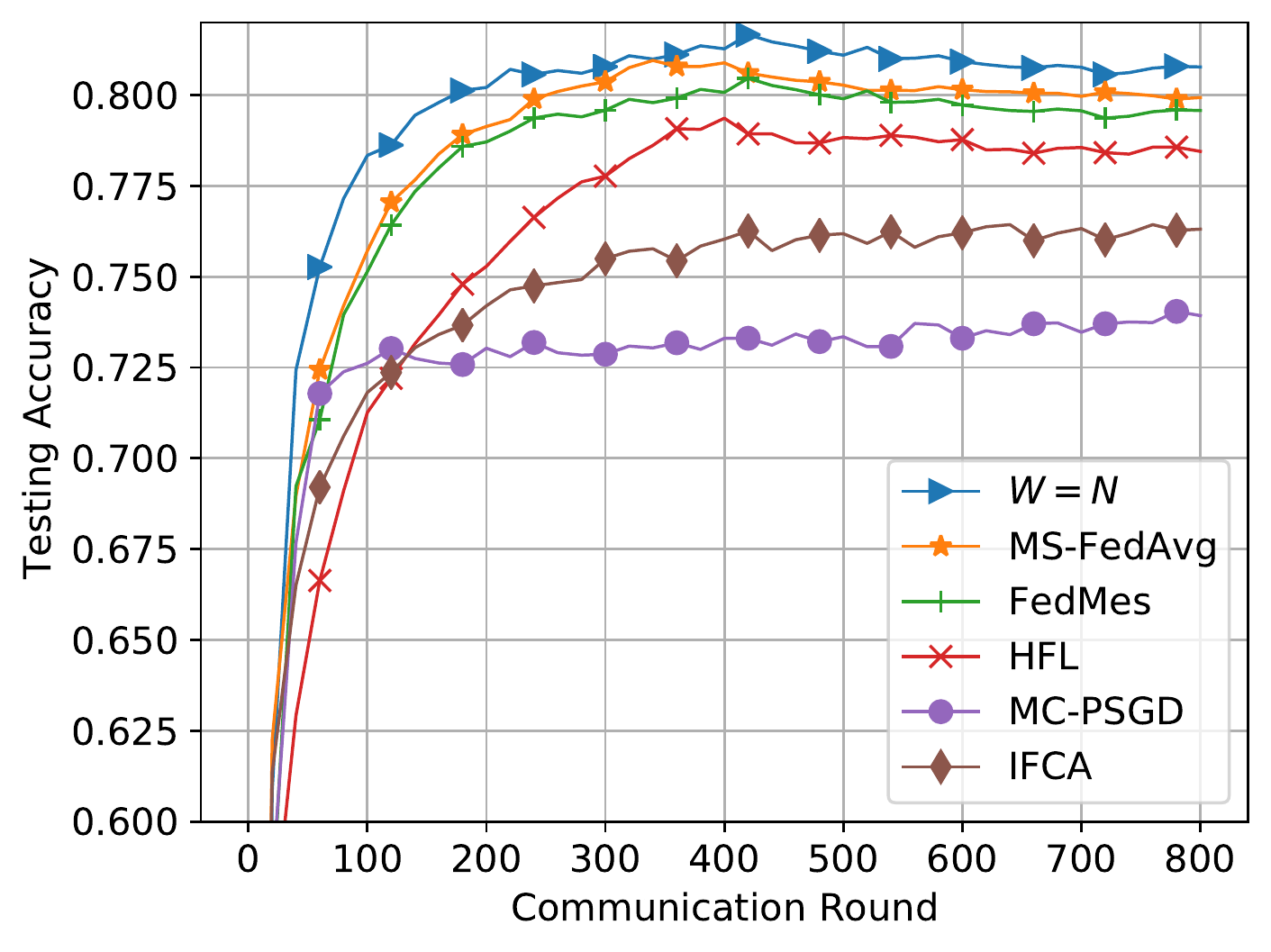}
\subcaption[second]{CIFAR-10 dataset.}
\end{minipage}%
\hfill
\begin{minipage}{0.64\columnwidth}
  \centering
\includegraphics[width=\textwidth, height=0.7\columnwidth]{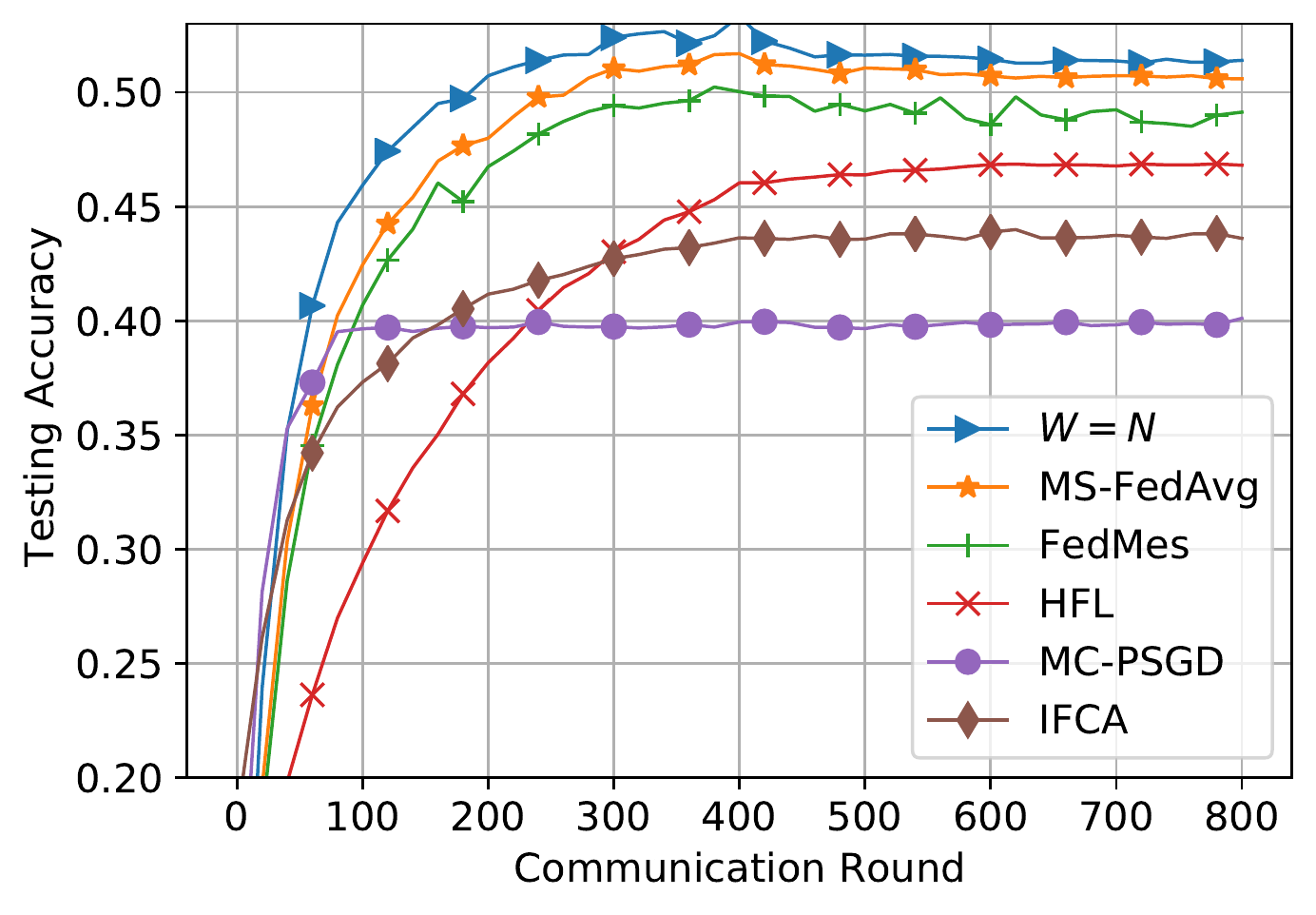}
\subcaption[third]{CIFAR-100 dataset.}
\end{minipage}
\caption{Testing accuracy for full client participation on multi-server FL, HFL, and CFL architectures.}
\label{Fig:fullclient}
\end{figure*}

\begin{figure*}[t!]
\centering
\begin{minipage}{0.64\columnwidth}
  \centering
\includegraphics[width=\textwidth, height=0.7\columnwidth]{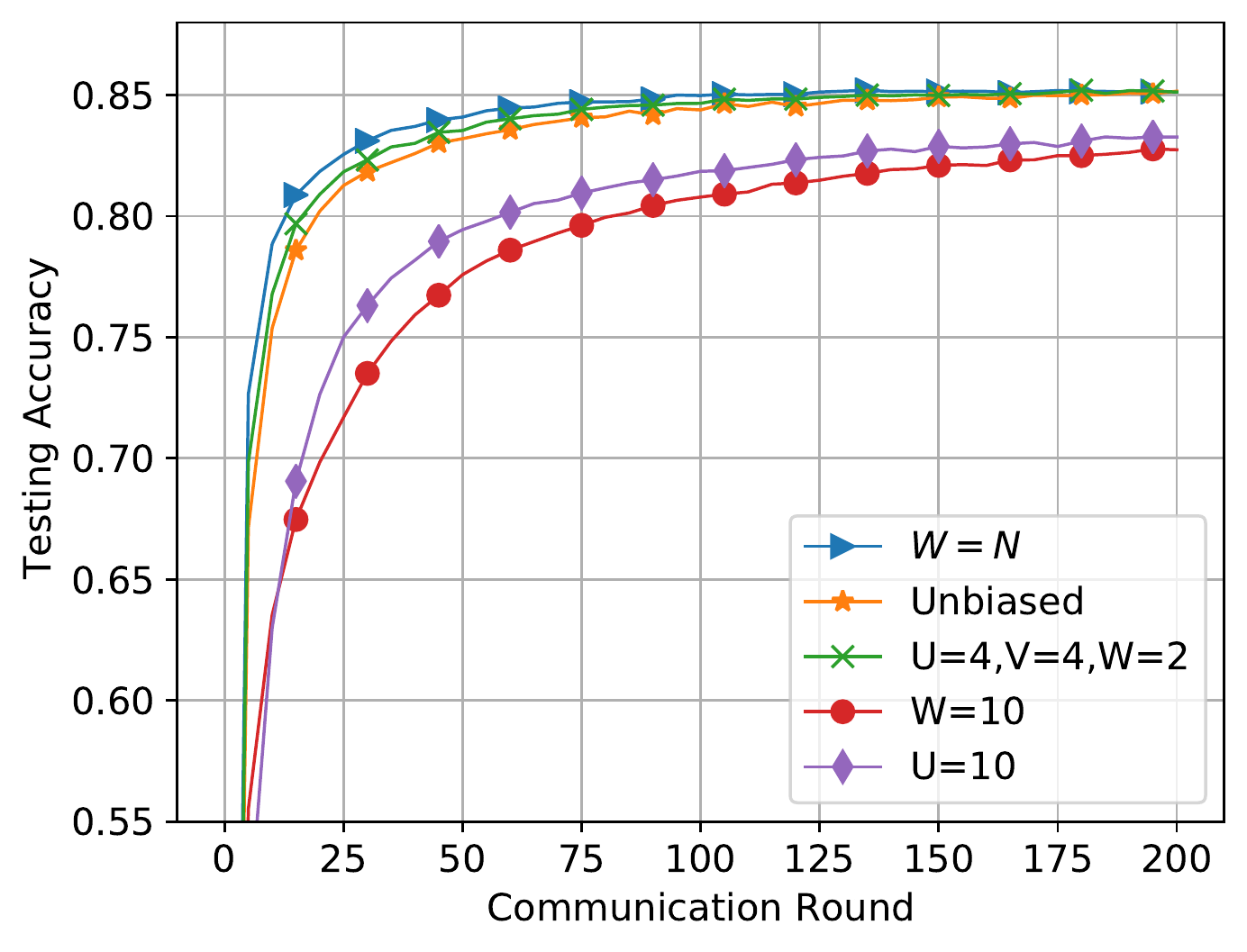}
\subcaption[first]{EMNIST dataset.}
\end{minipage}
\hfill
\begin{minipage}{0.64\columnwidth}
  \centering
\includegraphics[width=\textwidth, height=0.7\columnwidth]{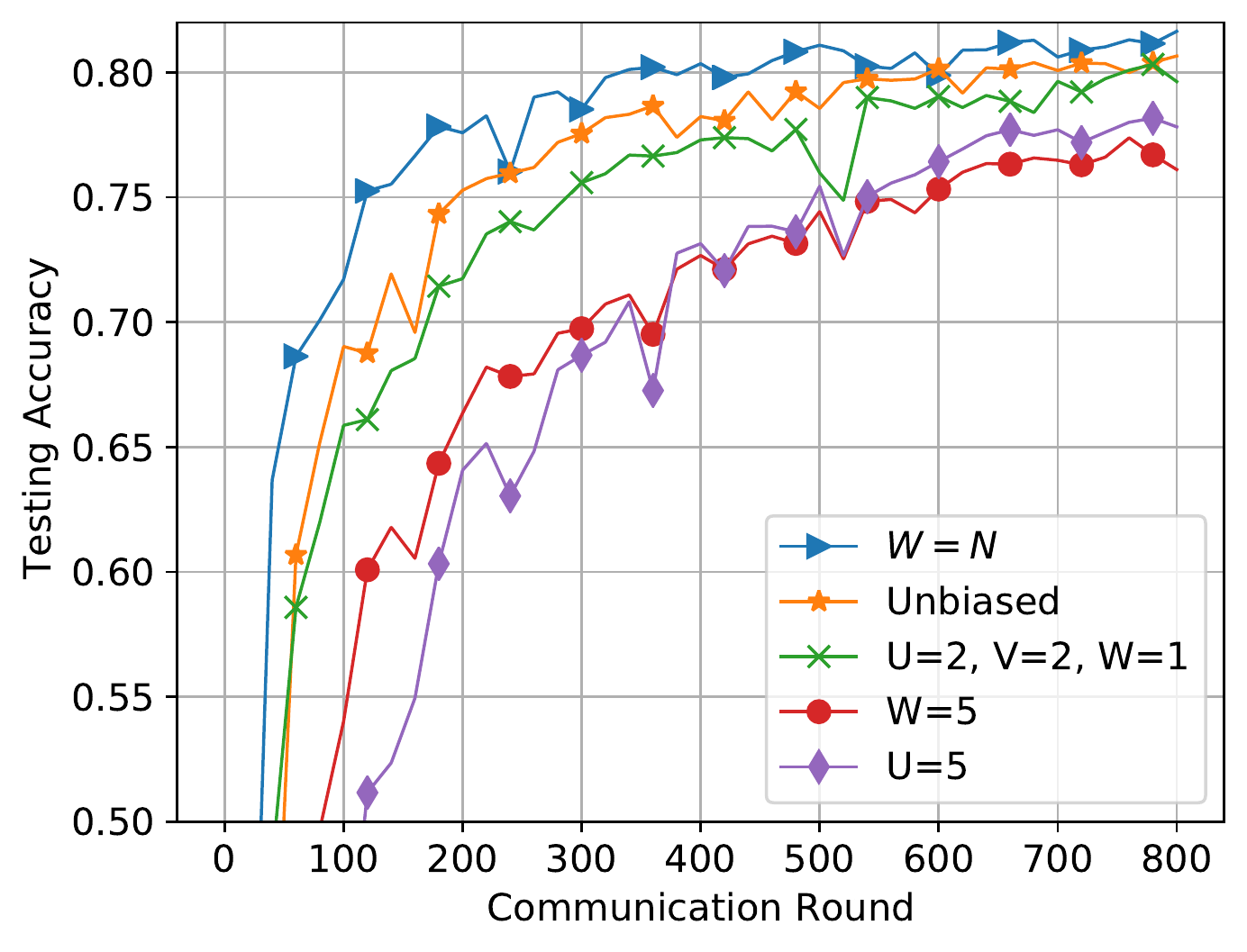}
\subcaption[second]{CIFAR-10 dataset.}
\end{minipage}%
\hfill
\begin{minipage}{0.64\columnwidth}
  \centering
\includegraphics[width=\textwidth, height=0.7\columnwidth]{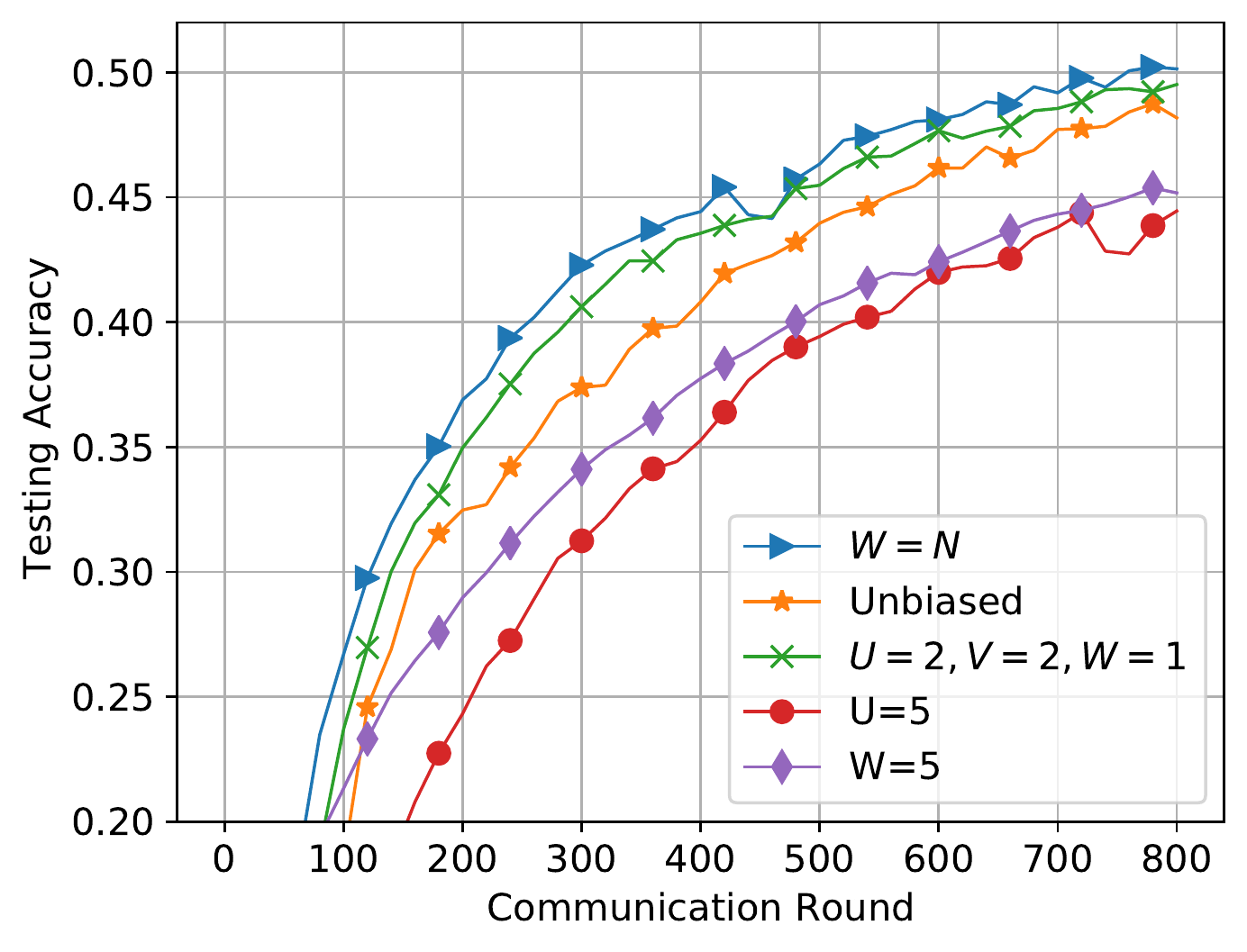}
\subcaption[third]{CIFAR-100 dataset.}
\end{minipage}
\caption{Testing accuracy for partial client participation on MS-FedAvg.}
\label{Fig:partialpart}
\end{figure*}

\textbf{Compared benchmarks.} In this paper, we compare our proposed algorithms to 5 existing FL benchmarks and can be concluded into 3 categories, i.e., single-server FL, HFL and clustered FL. 

\textbf{(1) FedAvg:} FedAvg algorithm \cite{mcmahan2017communication} is the most important baseline in FL research field. Note that the setting of FedAvg is same as \cite{yang2021achieving}.

\textbf{(2) Fedprox:} Fedprox \cite{li2018federated} develops a $l_2$-norm regularized algorithm to address the local model updates in the heterogeneous FL. In our experiment, we follow the settings in \cite{li2018federated} with $\lambda = 0.01$, which controls the dissimilarity of local objectives.

\textbf{(3) HFL:} HFL \cite{liu2021hierarchical, wang2020local} is a edge-cloud based FL architecture. In our experiment, we use one layer edge servers, and after 5 times client to edge server communication rounds, edge servers upload the model to the cloud server to compute aggregation.

\textbf{(4) MC-PSGD:} MC-PSGD \cite{ding2020distributed} is a CFL architecture, which processes the local training by clustering the clients into several blocks to reduce the client drift. We assume that re-clustering the blocks in each communication round, and the re-clustering time $\tau_{\text{Cluster}}$ is $1/20$ of one round.

\textbf{(5) IFCA:} IFCA \cite{ghosh2020efficient} is a clustered FL, which is clustered every 5 times communication round and based on calculating the cosine similarity, where $\tau_{\text{Cluster}}$ is $1/20$ of one communication round.

{\hlb \textbf{(6) FedMes:} FedMes \cite{han2021fedmes} is a multi-server FL, which sets the $\eta_g = 1$.}

\begin{table*}[t!]
\centering
\caption{Final testing accuracy, round and wall-clock(sec) with compared benchmarks: 80\% for EMNIST dataset, 75\% for CIFAR-100 dataset, and 45\% for CIFAR-100 dataset.}
\adjustbox{width=1.8\columnwidth}{\begin{tabular}{cccccccccc}
\hline
Dataset & \multicolumn{3}{c}{EMNIST} & \multicolumn{3}{c}{CIFAR-10} & \multicolumn{3}{c}{CIFAR-100} \\ \hline 
       Algorithm                      & Accuracy  & Round  & Wall-clock  & Accuracy  & Round  & Wall-clock & Accuracy & Round & Wall-clock \\ \hline
FedAvg                      & 85.06\%      & 8    & 5.37   & 80.75\%      & 88    & 1495.21  & 50.25\%      & 95    & 2119.45\\ 
FedProx                           & 84.97\%      & 10    & 6.72 & 80.06\%      & 106    & 1803.06  & 49.61\%      & 110    & 2643.30\\ 
HFL                         & 84.87\%      & 31    & 16.12    & 77.52\%      & 191     & 2555.58   & 46.19\%      & 351    & 7563.98  \\ 
MC-PSGD                         & 83.03\%      & 68    & 36.04    & 73.86\%      & NA     & NA   & NA      & NA    & NA  \\ 
IFCA                       & 83.85\%      & 65    & 34.13    & 76.61\%      & 291    & 4367.93 & NA      & NA    & NA \\
{\hlb FedMes}                        & {\hlb 84.91\%}      & {\hlb 16}    & {\hlb 7.64}    & {\hlb 79.08\%}      & {\hlb 100}    & {\hlb 962.17}  & {\hlb 49.82\%}      & {\hlb 130}    & {\hlb 2056.71  } \\
$W=N$                        & 85.04\%      & 11    & 5.97    & 80.98\%      & 85    & 785.90  & 50.62\%      & 90    & 1632.62   \\
MS-FedAvg                      & 85.02\%      & 13    & 7.15    & 79.84\%      & 91    & 903.17   & 50.14\%      & 119    & 1959.72  \\\hline
\end{tabular}}
\label{Tab:clientnumber}
\end{table*}

\textbf{Multi-server network architecture.} We set our multi-server FL network architecture with $M = 3$ regional servers and $85$ clients. Here, we consider a symmetric geometry multiple servers network with $U = 15$, $V = 10$ and $W = 10$ ($U$ is the number of clients in the non-overlapping area for every server, $V$ is the number of clients in the overlapping area between any two servers, $W$ is the number of clients in the overlapping area among all three servers) such that each regional server covers $45$ clients, which is shown in Fig.~\ref{Fig:symmetric}. Another multi-server FL network is that all $85$ clients are within in the overlapping area among all three servers and hence $W = 85$, $U = 0$ and $V = 0$. For the partial participation strategy, each regional server randomly samples 10 clients in each communication round. The asymmetric network architecture will be presented later.

\textbf{Network parameters setup.} The network setting is summarized as follows unless otherwise specified. We consider the regional server with a disc of 2km and cloud server with 5km. The channel gain of both the uplink and downlink are composed of both small-scale fading and large-scale fading. The small-scale fading is set as Rayleigh distribution with uniform variance and the large scale fading from client to regional server, client to cloud server and regional server to cloud server are all generated using the path-loss model $P_L = 128.1 + 37.6 \log_{10} (d(\text{km}))$, where $d$ is the distance in km. The noise power $\mu^2$ is $-107$ dBm. Total bandwidth budget $B_{rc} = 850$MHz, $B_{cr} = 475$MHz and $B_{cc} = 150$MHz. Both the uplink and downlink transmit power is $23$dBm, i.e., $p_i^U = p_i^D = 23$dBm, $\forall i \in \mathcal{N}$. {\hlb These parameters are followed by the existing edge computing studies \cite{tran2018joint, chen2019budget, farhadi2021service}.}

\subsection{Comparison to Other Multi-Server FL benchmarks}
In this subsection, we mainly focus on comparing the performance with the multi-server FL benchmarks, and including three settings: full clients participation, partial clients participation and moving clients scenarios. Note that the setting $W=85$ means that all 85 clients locate into the overlapping area with 3 regional servers, which is considered as the upper bound of MS-FedAvg, since all the three regional models reduce the divergence of the initial model.

\textbf{(1) Performance of full clients participation strategies:} In Figure~\ref{Fig:fullclient}, we aim to show the performance of full clients participation strategy compared to the three different multi-server FL benchmarks. It is easy to see that our proposed MS-FedAvg algorithm converges faster and achieves the best accuracy performance than other benchmarks in all the three datasets except for the setting with $W=85$, which supports our theoretical results in Theorem~\ref{The:nonconvexfull}. For example, in CIFAR-10 dataset, MS-FedAvg can achieve 79.69\% testing accuracy, which is 1.51\%, 3.78\% and 5.59\% higher than HFL, IFCA and MC-PSGD. In particular, MC-PSGD converges fast but it achieves lowest accuracy, since the clustered model is easy to overfit to its own cluster. However, the global model performance is worst among all benchmarks, i.e., 74.10\%. More specially, the disadvantage of HFL is due to that fact that for every 5 regional aggregation steps, it is required a global aggregation, which may degrade all the regional learning performance. {\hlb Although FedMes outperforms other three benchmarks, it does not achieve better convergence rate and accuracy than our proposed MS-FedAvg. The results may be due to the fact that the value of $\eta_g$ is not optimal.}

\begin{figure*}[t!]
\centering
\includegraphics[width=0.9\textwidth, height=0.26\textwidth]{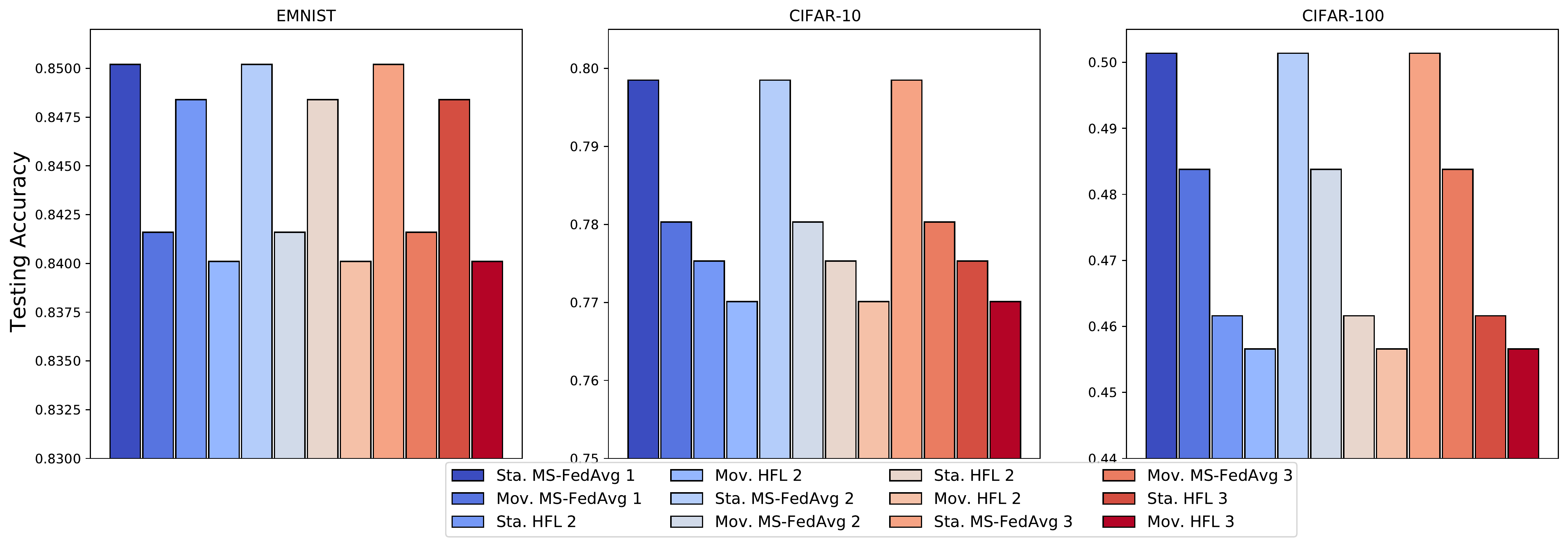}
\caption{Testing accuracy of static and moving scenarios.}
\label{Fig:moveacc}
\end{figure*}

\textbf{(2) Performance of partial clients participation strategies:} For the partial clients participation strategy, we uniformly sample $K=10$ clients in each region and communication round, and the performance of different values of $K$ will be shown later. For convenience, the meaning of legend is the number of sampled clients in each area types. $U$ is the isolated area type, $V$ is the clients located in the overlapping area with two regional servers and $W$ is the three regional servers' ovelapping area. For example, $U=10$ is sampling 10 clients in area type $U$. Based on this network architecture, we have the following interesting observations. 

\begin{figure*}[t!]
\centering
\includegraphics[width=0.9\textwidth, height=0.26\textwidth]{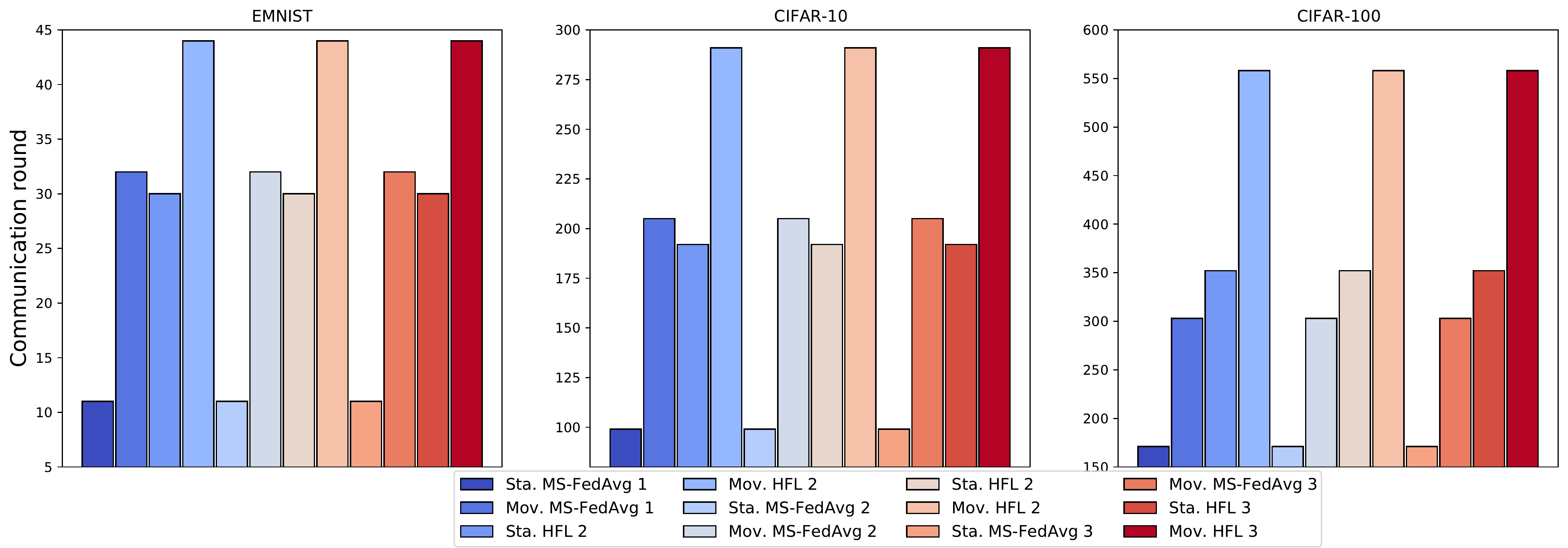}
\caption{Communication round to achieve the targeted accuracy of static and moving scenarios.}
\label{Fig:moveround}
\end{figure*}

Firstly, we can see that the unbiased partial participation of all the three datasets in Fig.~\ref{Fig:partialpart} is similar to the performance of full client participation in Fig.~\ref{Fig:fullclient} but higher variance, which is due to the uniformly sampling, and successfully matches our analysis in Section~\ref{Sec:Discussion}. If we select these 10 clients with the number of $U=4, V=4$ and $W=2$, it performs better than unbiased participation strategy, e.g., 1.43\% higher testing accuracy than unbiased MS-FedAvg in CIFAR-100 dataset. More specifically, if we only sample clients in one specific area, the performance has much degradation, e.g., 45.01\% with $W=10$, and 43.92\% with $U=10$. The reason may be due to the fact that that the regional model overfits these local clients and cannot generalize to all clients in the entire FL network. Therefore, if we use biased participation strategy, it is necessary to sample clients among all area types. The learning performance of Biased MS-FedAvg strongly depends on the network topology, and hence it is not easy to provide the optimized sampling strategy. However, it is feasible to find a sampling strategy which performs better than unbiased strategy.

\textbf{(3) Performance of the clients movement scenarios:} Here, we aim to show the comparison of static and movement scenarios of multi-server FL settings. Because clustered FL needs to re-cluster every several communication rounds, the movement scenario can be ignored in this setting. Therefore, we only compare our proposed MS-FedAvg algorithm to HFL. Since we cannot justify the deterministic moving direction of each client, we assume that it randomly moving in each communication round. {\hlb Because of the restricted service area of regional server, we consider the movement scenario such as the moving sensors or IoT devices \cite{wang2020edge, kuang2017energy}, which performs low movement speed (e.g., 3 miles per hour \cite{kuang2017energy}). Compared to the transmission latency, the moving distance is very short and we can assume that each client connect the same corresponding regional server(s) within one communication round. In order to evaluate the training performance between static and movement scenarios, we set three network settings: (1) the probabilities of the client locating in each area are $\mathbb{P}(\text{locate~in}~U) = 52.94\%$, $\mathbb{P}(\text{locate~in}~V) = 35.30\%$ and $\mathbb{P}(\text{locate~in}~W) = 11.76\%$; (2) $\mathbb{P}(\text{locate~in}~U) = 35.30\%$, $\mathbb{P}(\text{locate~in}~V) = 52.94\%$ and $\mathbb{P}(\text{locate~in}~W) = 11.76\%$; and (3) $\mathbb{P}(\text{locate~in}~U) = 52.94\%$, $\mathbb{P}(\text{locate~in}~V) = 11.76\%$ and $\mathbb{P}(\text{locate~in}~W) = 35.30\%$.} The communication round is to achieve 80\% on EMNIST dataset, 75\% on CIFAR-10 dataset and 45\% on CIFAR-100 dataset.

Clearly, we can see that the convergence rate of movement scenarios is much slower than static scenarios among all the multi-server FL settings, e.g., in CIFAR-10 dataset, movement is 77.08\% and static is 79.84\% of MS-FedAvg. And it only uses 91 communication rounds to achieve 75\% testing accuracy, which is much better than movement scenario with 207 rounds. The reason is that since the clients may participate in different regional models training, it incur higher model variance between each communication round. As a result, it makes the global model to be converged difficultly. It is similar to train the same several regional models on each regional server. Therefore, it is not necessary to consider the movement scenario in this paper. In addition, it is clear to see that our proposed MS-FedAvg also outperforms other benchmarks in the movement scenarios. {\hlb If we assume more clients locate in the overlapping areas (e.g., setting 3), the training performance of both MS-FedAvg and HFL improve. For example, in setting 3, the movement scenario of MS-FedAvg has 48.75\% testing accuracy and uses 279 to achieve 45\% accuracy on CIFAR-100 dataset. This may be because the data distribution clients performs less diversity and near to the $W=N$ scenario. More specifically, MS-FedAvg outperforms HFL in all settings (e.g., on CIFAR-10 dataset of setting 2, the movement scenario of MS-FedAvg achieves 78.01\% and HFL is 76.18\%).}

\subsection{Training Performance and Transmission Latency}
In this subsection, since it is not easy to verify the local computing time of each client, and the existing papers have shown that \cite{kairouz2019advances, mcmahan2017communication} the transmission latency dominate the running time FL, and hence we only compare the transmission latency to training performance of all FL benchmarks and simply ignore the local computing time of every client. The wall-clock means that the total transmission time to achieve the targeted testing accuracy.

Table~\ref{Tab:clientnumber} shows the final testing accuracy, communication round and wall-clock to achieve the targeted testing accuracy. We compare our MS-FedAvg algorithm to single-server FL, HFL and CFL. It is easy to observe that the final testing accuracy of $W=N$ and FedAvg perform similar among all the three datasets, e.g.,. 78.98\% and 78.96\% in CIFAR-10 dataset. This is from that they do not have the model divergence to degrade the learning performance. The $W=N$ setting can be considered as the FedAvg on multi-server FL network architecture, while $W=N$ is much more efficient from the transmission latency perspectives. For the EMNIST dataset, the reason that FedAvg algorithm has the best performance is due to the fact that EMNIST dataset is simple, and easy to achieve targeted testing accuracy. For the more complicated datasets, it is clearly to see that MS-FedAvg outperforms single-server FL benchmarks. {\hlb Note that FedMes outperforms other three benchmarks, but it is worse than MS-FedAvg.}

\begin{figure*}[t!]
    \centering
    \begin{subfigure}{0.66\columnwidth}
    \centering
    \includegraphics[width=0.48\columnwidth]{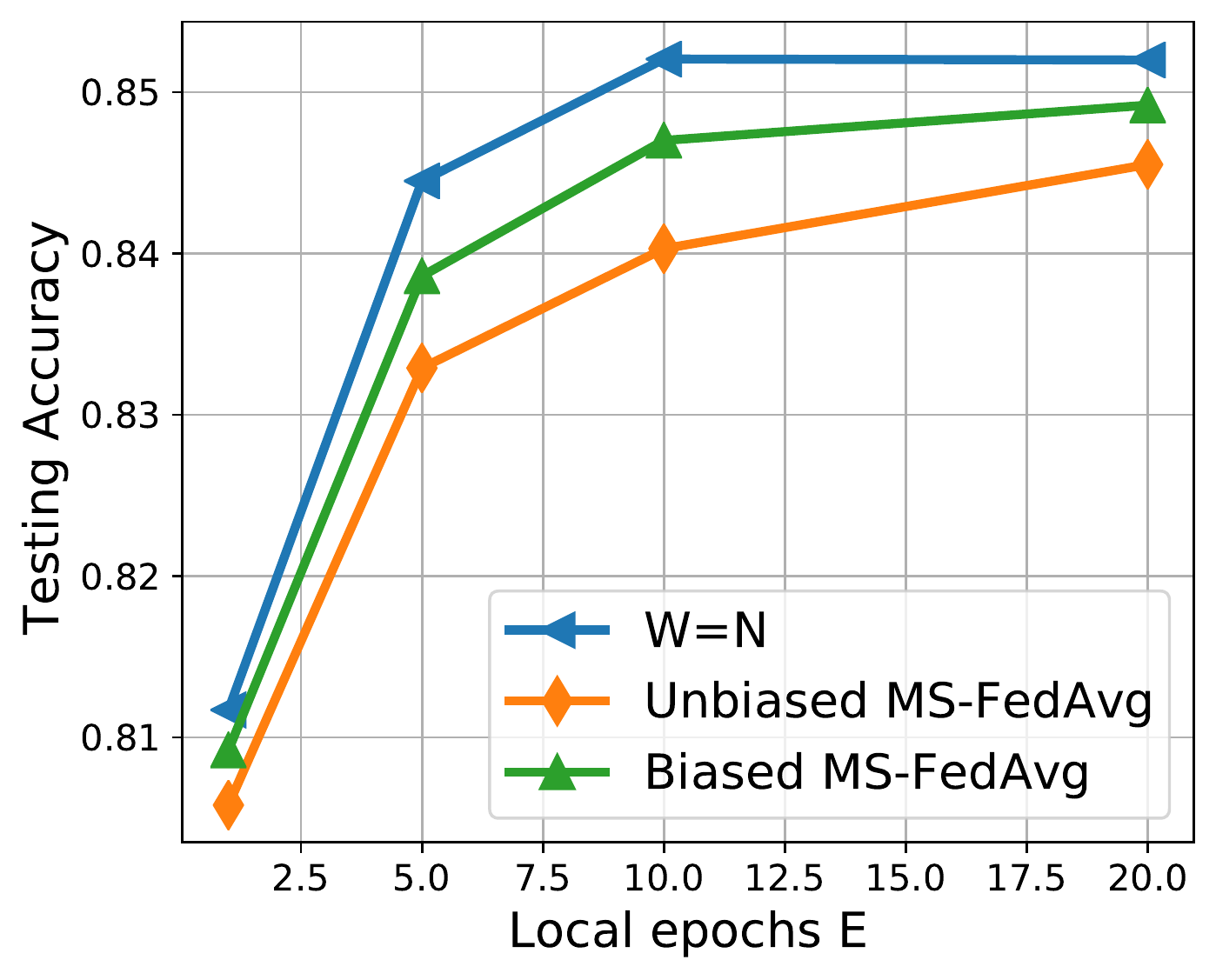}
    \includegraphics[width=0.48\columnwidth]{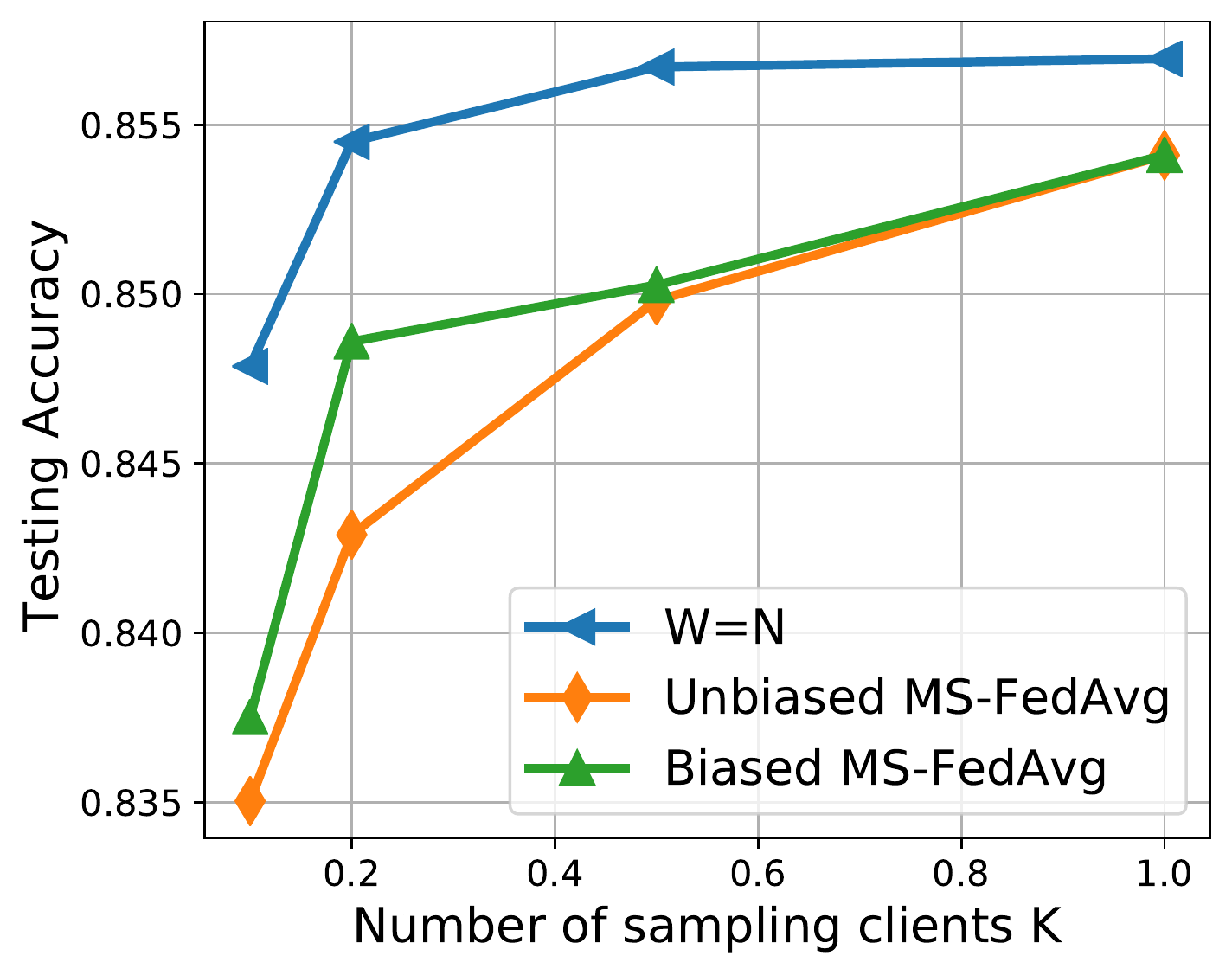}
    \subcaption{EMNIST dataset.}
    \label{Fig:impactemnist}
    \end{subfigure}
    \begin{subfigure}{0.66\columnwidth}
    \centering
    \includegraphics[width=0.48\columnwidth]{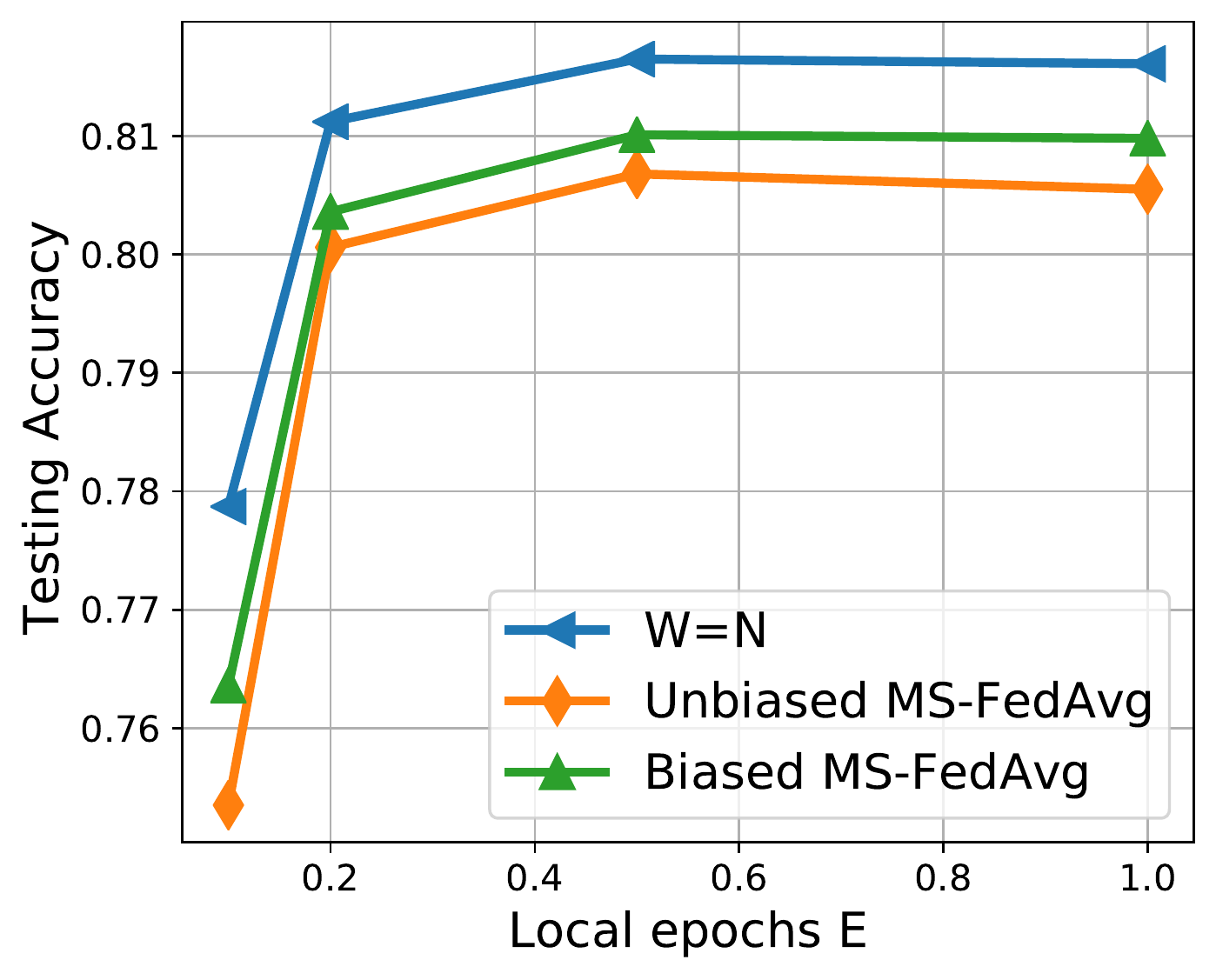}
    \includegraphics[width=0.48\columnwidth]{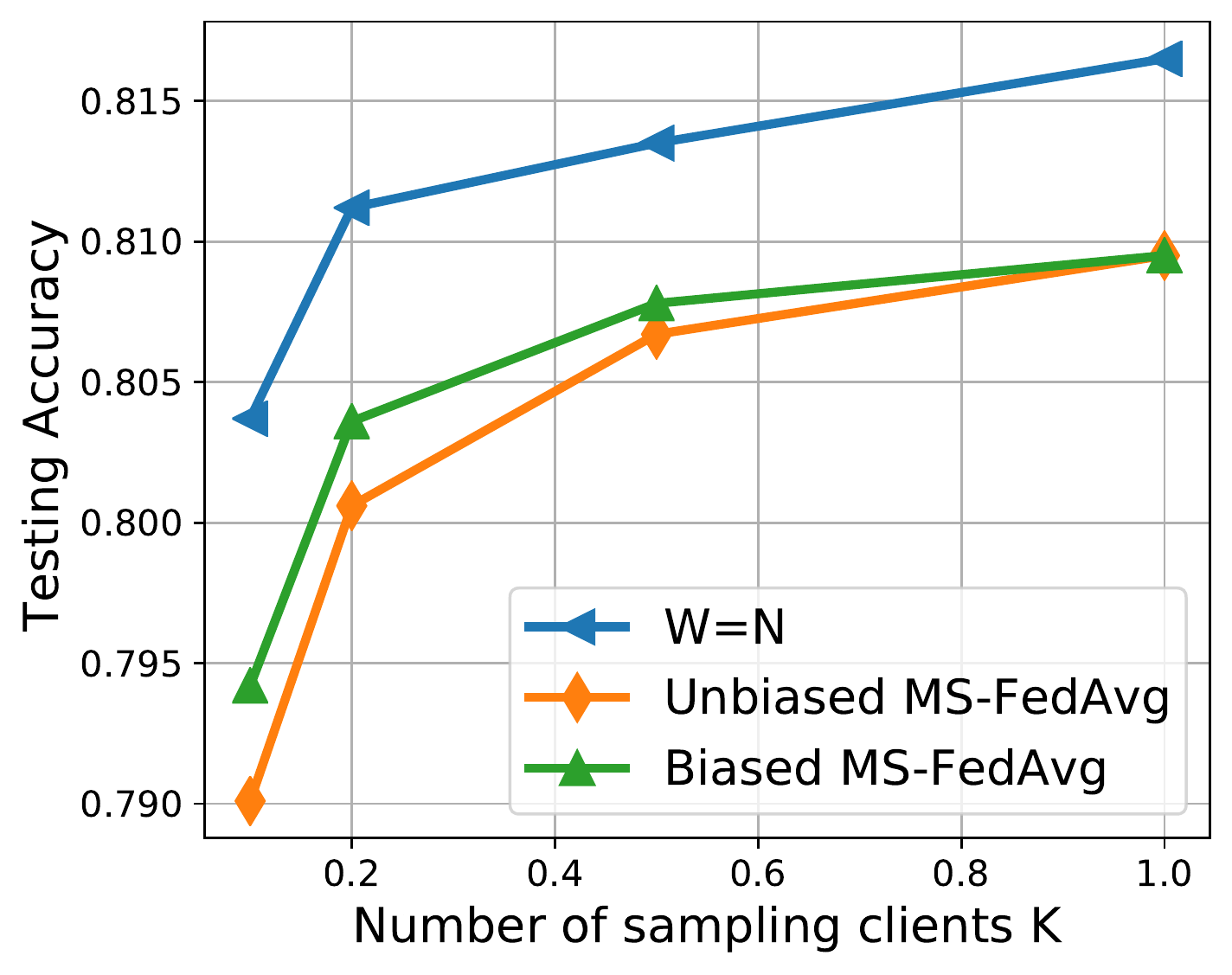}
    \subcaption{CIFAR-10 dataset.}
    \label{Fig:impactcifa10}
    \end{subfigure}
    \begin{subfigure}{0.66\columnwidth}
    \centering
    \includegraphics[width=0.48\columnwidth]{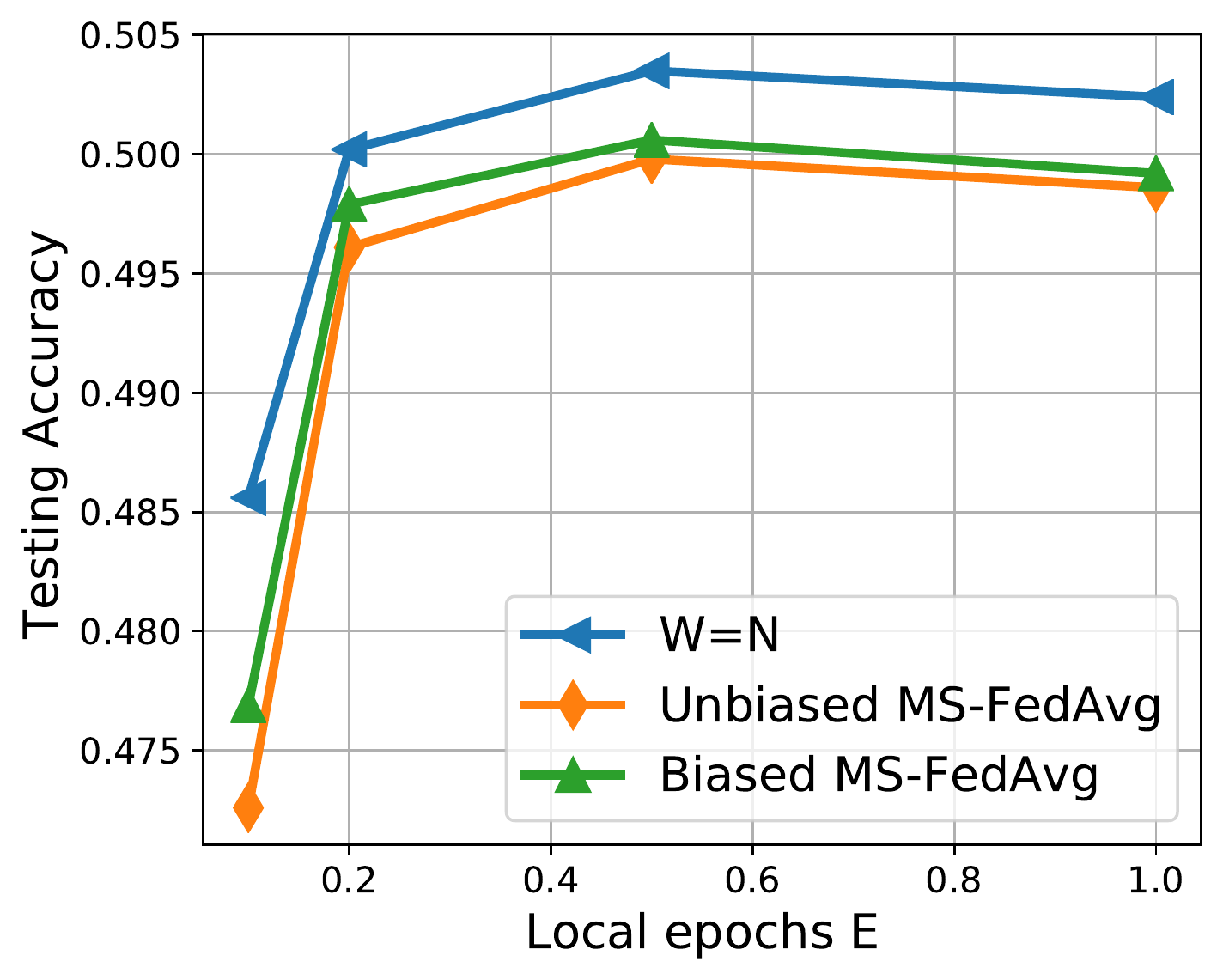}
    \includegraphics[width=0.48\columnwidth]{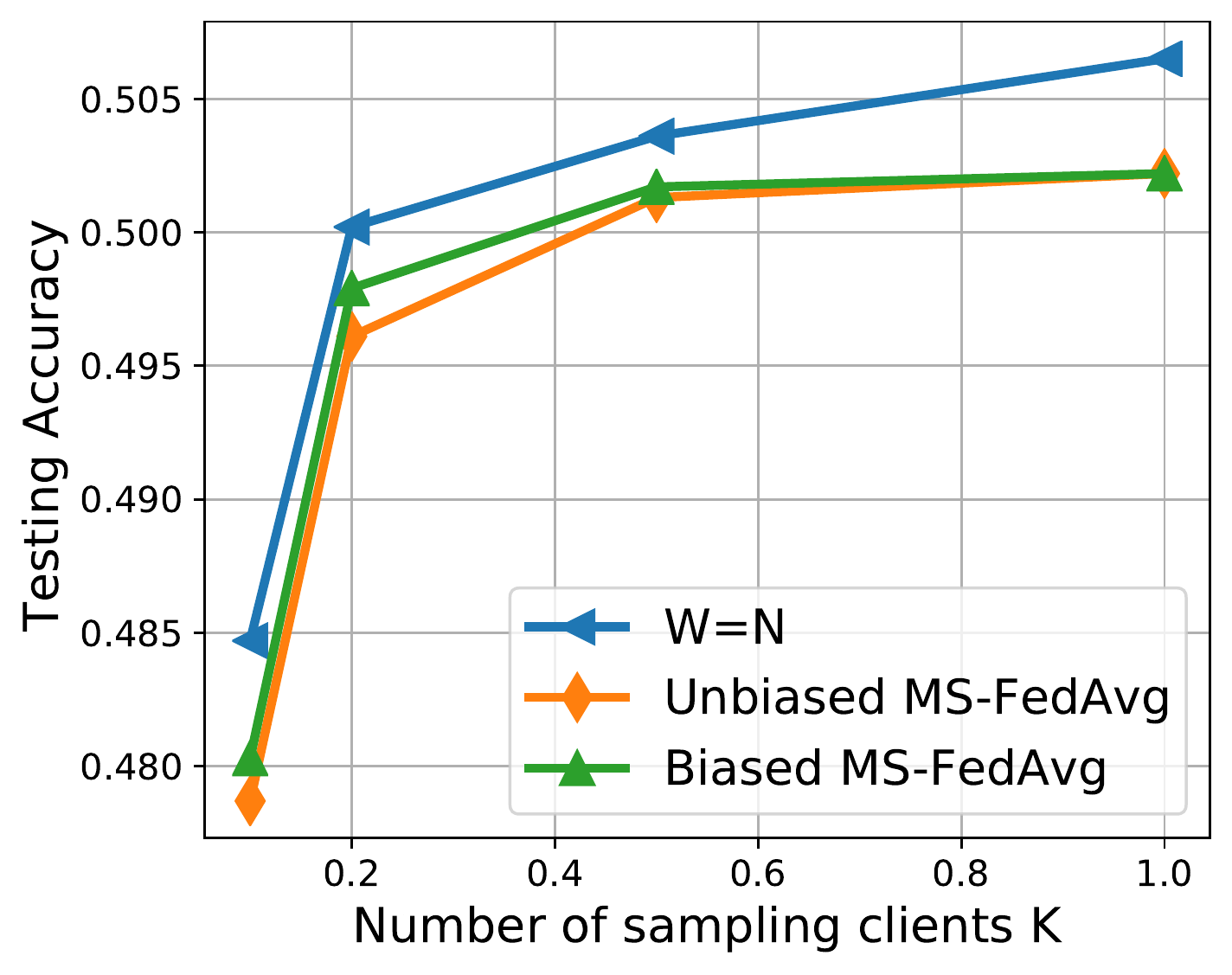}
    \subcaption{CIFAR-100 dataset.}
    \label{Fig:impactcifar100}
    \end{subfigure}
    \caption{Impact on the number of sampling clients $K_m$ and the number of local epochs $E$.}
    \label{Fig:impact}
\end{figure*}

\begin{table*}[t!]
\centering
\caption{{\hlb Impact on different bandwidth settings.}}
\adjustbox{width=1.75\columnwidth}{\begin{tabular}{ccccccc}
\hline
{\hlb Dataset} & \multicolumn{2}{c}{{\hlb EMNIST}} & \multicolumn{2}{c}{{\hlb CIFAR-10}} & \multicolumn{2}{c}{{\hlb CIFAR-100}} \\ \hline
{\hlb Bandwidth} & {\hlb Accuracy}  & {\hlb Wall-clock}  & {\hlb Accuracy}  & {\hlb Wall-clock} & {\hlb Accuracy} & {\hlb Wall-clock} \\ \hline
{\hlb $b_{rc} = 10$MHz, $b_{cr} = 5$MHz} & {\hlb 85.02\%} & {\hlb 7.15}  & {\hlb 79.84\%} & {\hlb 903.17} & {\hlb 50.14\%} & {\hlb 1959.72} \\ 
{\hlb $b_{rc} \sim \mathcal{U} [8, 12]$MHz, $b_{cr} \sim \mathcal{U}[4,6]$MHz}   & {\hlb 85.05\%} & {\hlb 10.49} & {\hlb 79.59\%}  & {\hlb 1003.26}   & {\hlb 49.98\%} & {\hlb 2227.01} \\ 
{\hlb $b_{rc} \sim \mathcal{U} [5, 15]$MHz, $b_{cr} \sim \mathcal{U}[2,8]$MHz}   & {\hlb 84.97\%}   &{\hlb 15.01} &{\hlb 79.86\%} & {\hlb 1420.39}   & {\hlb 50.05\%} & {\hlb 2936.91} \\ \hline
\end{tabular}}
\label{Tab:Bandwidth}
\end{table*}

Although both the two single-server FL benchmarks perform good accuracy performance, they will waste much more training time, due to the large averaged distance between clients and server. Although the HFL and CFL algorithms spend less transmission latency for one communication round, it waste much more wall-clock time to achieve the targeted accuracy due to the low convergence rate, e.g., for CIFAR-10 dataset, HFL uses 191 round and 2555.58 sec, and IFCA uses 291 rounds and 4367.93 sec. Therefore, the existing multi-server FL benchmarks cannot guarantee to be efficient enough from the transmission perspective.

Our proposed MS-FedAvg outperforms other multi-server FL benchmarks on testing accuracy perspectives. More specifically, MS-FedAvg has the best wall clock-time among all the benchmarks except the $W=N$ setting, since it does not need to download and upload models to the central servers, it significantly reduce the distance and save much more transmission latency. For example, in CIFAR-10 dataset, it saves 1.65$\times$, 1.99$\times$, 2.83$\times$ and 4.84$\times$ time than FedAvg, FedProx, HFL and IFCA. More specifically, due to the limited generalization of MC-PSGD, it cannot achieve the targeted accuracy. Therefore, our proposed MS-FedAvg algorithm can be considered as an efficient solution to address the bottleneck problem of FL settings.

\subsection{Impact on Different Parameters}
\textbf{(1) Impact on $K_m$ and $E$:} Based on our analysis in Sections~\ref{Sec:ConvergenceMSfedavg} and \ref{Sec:Discussion}, the learning performance of MS-FedAvg algorithm depends on several hyper-parameters, e.g., the number of sampling clients under each regional servers $K_m$ and the setting of number of local epochs $E$. Figs.~\ref{Fig:impactemnist}-\ref{Fig:impactcifar100} present the final testing accuracy of EMNIST, CIFAR-10 and CIFAR-100 datasets under different values of $K_m$ and $E$. Especially, we set $2U=2V=W$ as "Biased", which means the fraction of the number of sampling clients in different area types.

The results in Figs.~\ref{Fig:impactemnist}-\ref{Fig:impactcifar100} indicate that that the performance substantially improves when we increase the number of sampled clients number $K_m$, and the biased participation strategy consistently outperforms unbiased participation, e.g., in CIFAR-10 dataset, biased client participation strategy increases from 76.20\% to 82.93\%, when $K_m = 10$ and $35$, and unbiased increases from 75.56\% to 82.93\%. In addition, the degree of improvement of $K_m$ increases lower. This empirical result matches our analysis in Section~\ref{Sec:ConvergenceMSfedavg}, and performs similarly to single-server FL settings \cite{li2019convergence, yang2021achieving, karimireddy2020scaffold}.

Next, we aim to show the learning performance under different values of $E$. Until now, it is difficult to explicitly show the relationship between $E$ and learning performance. In \cite{li2019convergence}, they presented that increasing $E$ can improve the performance. However, other studies \cite{li2018federated, yang2021achieving} showed that when $E$ is set as too large, it will degrade the performance. Our experimental results in Figs.~\ref{Fig:impactemnist}-\ref{Fig:impactcifar100} imply that if $E=1$, it performs the worst. If we increase the value, the accuracy firstly increases but then decreases, e.g., in CIFAR-100, when $E=5$, the accuracy is 49.65\%, but 44.65\% of $E=20$. Thus, it is necessary to find a suitable value $E$ to achieve better performance on different datasets.

{\hlb \textbf{(2) Impact on bandwidth:} Here, we present the impact on different bandwidth settings between clients and regional server(s), which includes three settings: (1) $b_{rc} = 10$MHz, $b_{cr} = 5$MHz; (2) $b_{rc} \sim \mathcal{U} [8, 12]$MHz, $b_{cr} \sim \mathcal{U}[4,6]$MHz; and (3) $b_{rc} \sim \mathcal{U} [5, 15]$MHz, $b_{cr} \sim \mathcal{U}[2,8]$MHz, where $\mathcal{U}$ is uniform distribution.  

In Table~\ref{Tab:Bandwidth}, we can clearly see that different bandwidth does not have significant impact on the testing accuracy. For example, on EMNIST dataset, the testing accuracy of these three settings are 85.02\%, 85.05\% and 84.97\%. The is due to the fact that the learning performance is independent on the network parameter settings, and only depends on the setting of learning models (e.g., data distribution and hyper-parameters). However, the bandwidth has a large influence on the transmission latency, because each regional server should wait for the slowest client that performs small bandwidth and then process the aggregation. On CIFAR-100 dataset, if the bandwidth follows $b_{rc} \sim \mathcal{U} [5, 15]$MHz, $b_{cr} \sim \mathcal{U}[2,8]$MHz, the total communication time is 2936.91sec, which is 49.86\% higher than equal bandwidth setting. Therefore, if each regional server has limited bandwidth budget, the best way is to equally divided to each client, which can achieve the best performance on communication.}

\begin{figure}[t!]
    \centering
    \includegraphics[width=\FigWidthSmaller]{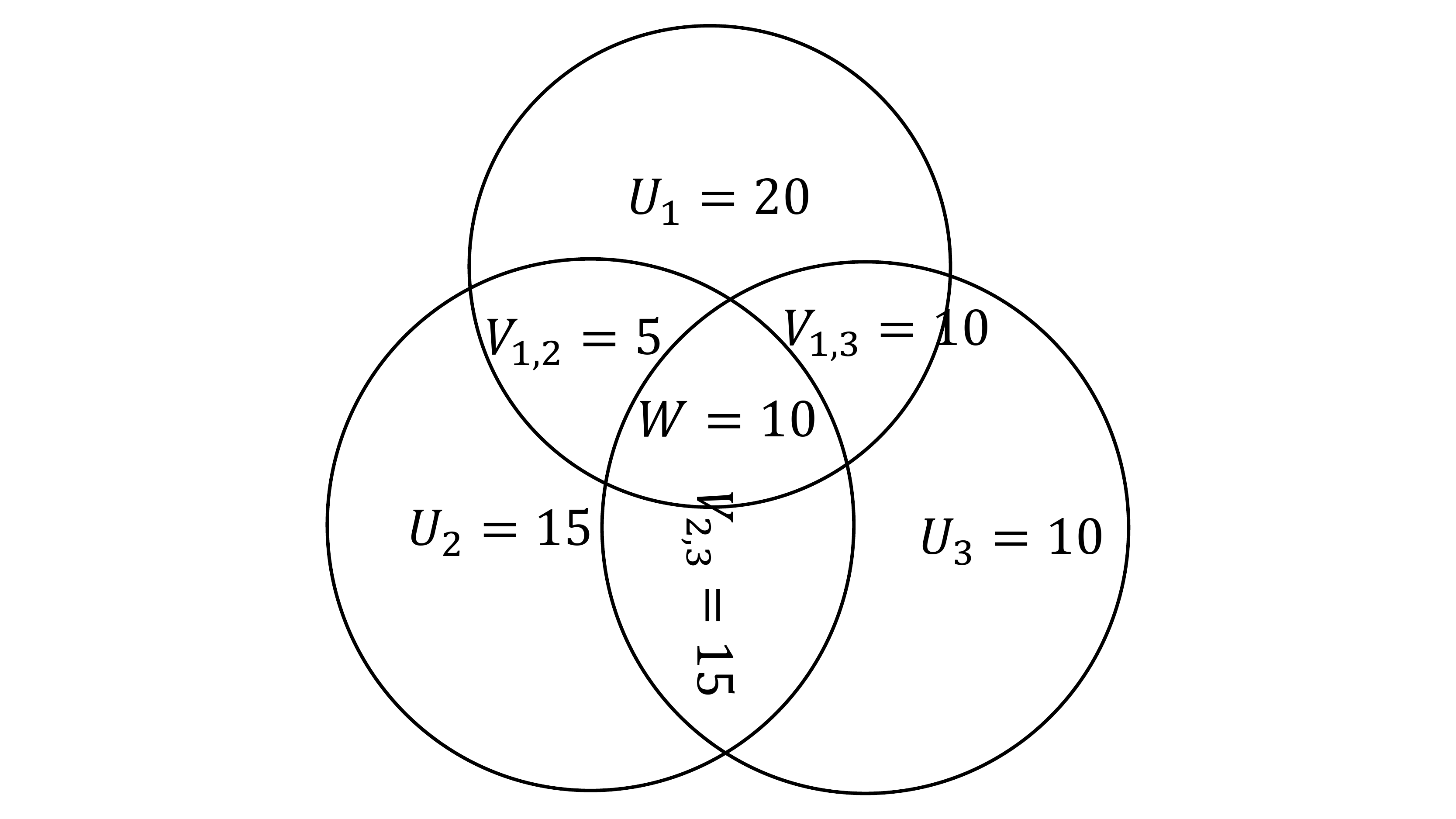}
    \caption{Asymmetric multi-server FL architecture.}
    \label{Fig:asymmetric}
\end{figure}

\begin{figure*}[t!]
\centering
\begin{minipage}{0.64\columnwidth}
  \centering
\includegraphics[width=\textwidth, height=0.7\columnwidth]{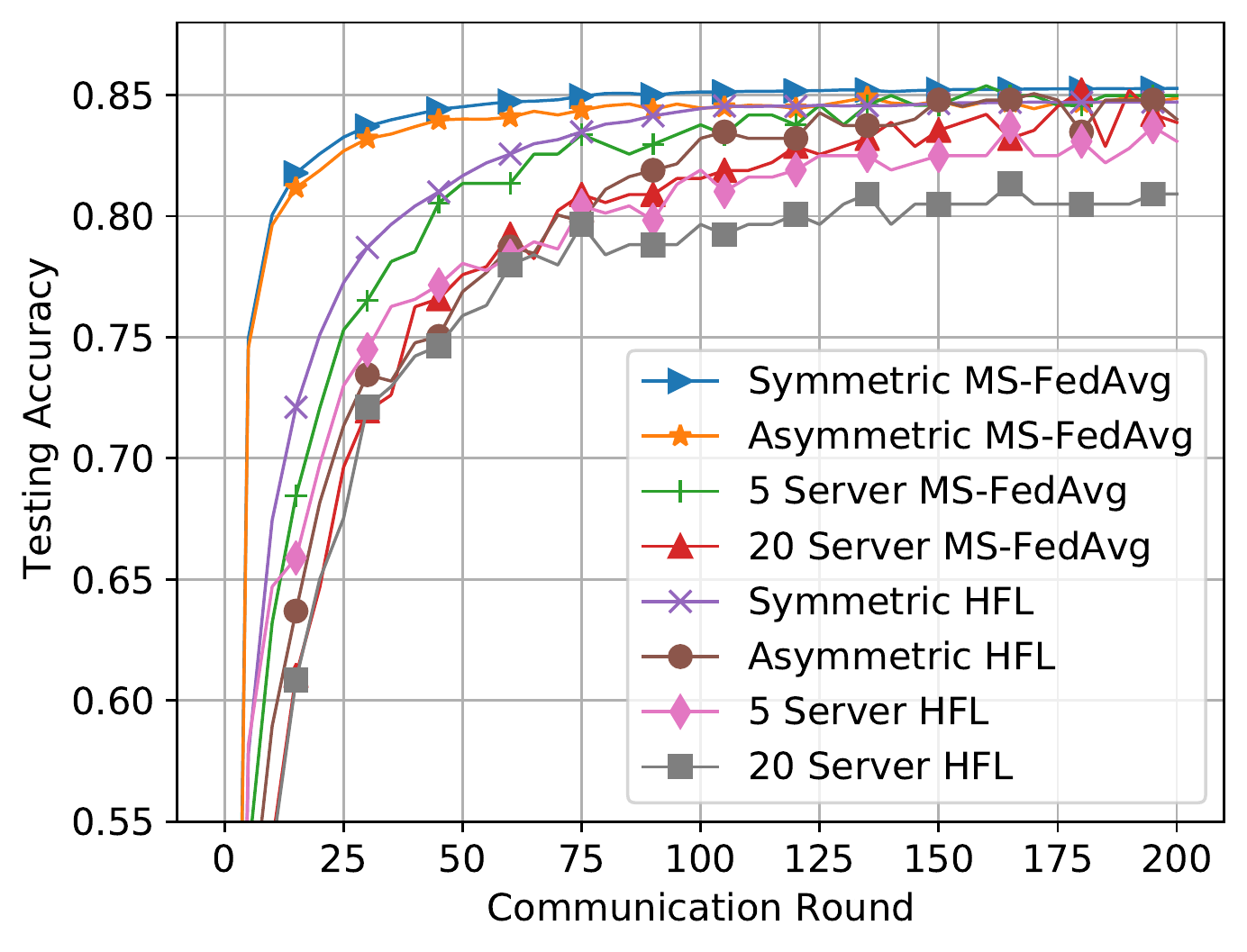}
\subcaption[first]{EMNIST dataset.}
\end{minipage}
\hfill
\begin{minipage}{0.64\columnwidth}
  \centering
\includegraphics[width=\textwidth, height=0.7\columnwidth]{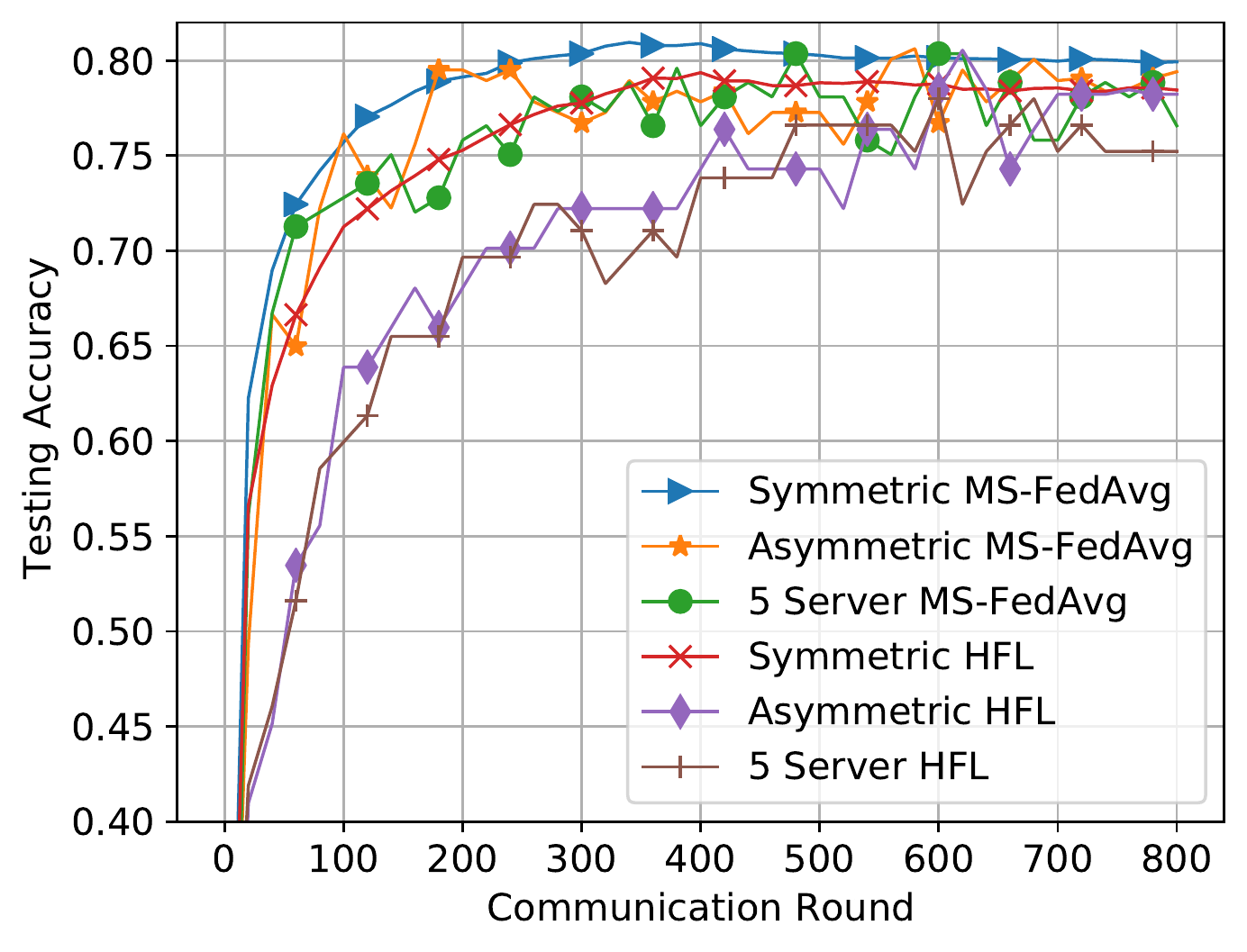}
\subcaption[second]{CIFAR-10 dataset.}
\end{minipage}%
\hfill
\begin{minipage}{0.64\columnwidth}
  \centering
\includegraphics[width=\textwidth, height=0.7\columnwidth]{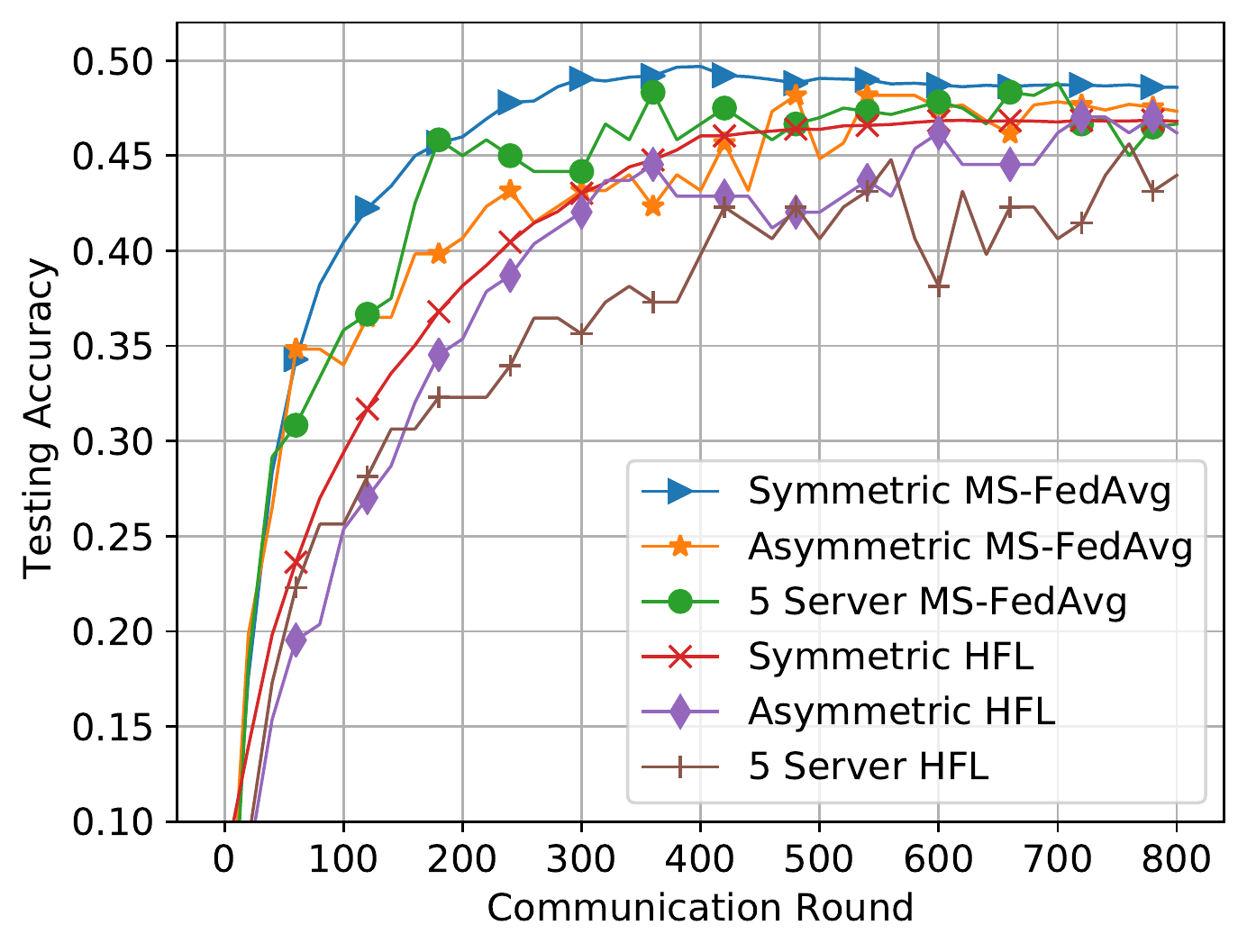}
\subcaption[third]{CIFAR-100 dataset.}
\end{minipage}
\caption{Impact on different multi-server FL network architectures.}
\label{Fig:differentnetwork}
\end{figure*}

\subsection{Impact on different Multi-Server FL Network Architectures}
Here, we aim to show the impact on multi-server FL network architecture. In this subsection, we additionally consider two more network architectures: asymmetric and 5 regional servers. As such, we simulate an asymmetric FL network architecture in Fig.~\ref{Fig:asymmetric}, which also includes 3 regional servers and 85 clients in total. In addition, we present the learning performance of multi-server FL network architecture with 5 regional servers. To make the regional servers efficiently serve the clients, we assume that the overlapping areas at most includes 3 regional servers, and also keep the network within 85 clients.  

In Fig.~\ref{Fig:differentnetwork}, it can be observed that the symmetric network architecture achieves the best learning performance on convergence perspective. The more complex network architecture, i.e., asymmetric and more regional servers, degrades the learning performance, especially incurring higher variance in each communication round. The observation may come from the unbalanced regional aggregation. Although they achieves the similar testing accuracy in final, symmetric network uses fewest communication rounds to achieve the targeted testing accuracy. {\hlb In order to show the training performance for the large-scale multi-server FL network, we embed the EMNIST dataset on a 20 regional servers, which includes 150 clients. The MS-FedAvg also substantially outperforms HFL, which can achieve 84.68\% testing accuracy (80.86\% for HFL). It indicates that MS-FedAvg works stably on complicated networks.} Designing the relationship between the learning performance and network architectures will be our further work. More specially, it is worth noting that all the learning performance under different networks, MS-FedAvg significantly outperforms HFL.

\section{Related Work}\label{Sec:Related}
Current research on FL wireless networks improved the communication efficiency on the following aspects. (1) how to properly allocate resource to clients. \cite{xu2020client} designs how to properly select clients and how bandwidth is allocated among the selected clients in each communication round. \cite{shi2020device} jointly considers bandwidth allocation and clients scheduling problem. For bandwidth allocation sub-problem, they aim to allocate more bandwidth to the clients with worse channel conditions, and develops a greedy policy to solve the clients scheduling sub-problem. (2) deadline based FL architecture. \cite{qu2021context} develops clients selection algorithm for deadline based HFL via contextual combinatorial multi-armed bandits to improve the training performance. (3) physical layer quantization. For bandwidth reduction, \cite{amiri2020federated} sparsifies the gradient estimates of clients to accumulate error from previous communication rounds, and project the resultant sparse into a low-dimensional vector. In \cite{zheng2020design}, they clarify how to communicate between clients and the central server and evaluate the impact on the various quantization. In addition, they design the physical layer quantization both on uplink and downlink. They mainly minimize the communication latency by solving an optimization problem subject to the constraint of obtaining a good model. However, few of them propose the details of the convergence guarantee in their papers.

FL was first proposed in \cite{mcmahan2017communication}, where they proposed the FedAvg algorithm and showed the advantages empirically on different datasets and local dataset distribution settings. Followed by \cite{mcmahan2017communication}, the authors propose the strategy to address the communication bottleneck problem by increasing the local training epochs \cite{li2019convergence, yang2021achieving}. Specifically, this method is also a feasible solution to improve the convergence rate. Based on this method, some new algorithms are developed from different perspectives. \cite{karimireddy2020scaffold} adds a variant control variable to reduce the local model updates and global model due to the non-iid distribution of local datasets, and \cite{lin2020ensemble} and \cite{li2021stragglers} designs FL algorithm for asynchronized FL via Hessian approximation. \cite{reddi2020adaptive} designs server level momentum and extends the local SGD optimizer to AdaGrad, YOGI and ADAM, \cite{karimireddy2021breaking, xu2021fedcm, li2021fedlga} proposes client level momentum FL algorithm, and \cite{khanduri2021stem} shows the impact on local batch size of both sided level momentum FL. However, these algorithms are mainly developed on single-server FL. In this paper, different from the above existing works, we derive the convergence results of the typical multi-server FL in \cite{han2021fedmes} that obtains the impact on non-iid datasets and the initial local models, which is much more challenging.

Based on the highly efficient edge computing architecture, some studies focus on edge facilitated FL: HFL \cite{lim2020federated, liu2020client, wang2020local} and clustered FL \cite{xie2020multi, lee2020tornadoaggregate, duan2020self}. However, they also rely on communicating to the central server, large communication delay is difficult to be avoidable compared to our proposed multi-server FL. Another direction of distributed learning is fully decentralized/serverless \cite{li2021lomar, roy2019braintorrent, pmlr-v139-kong21a}. In decentralized FL, clients need to exchange the model updates with their neighbors not to the servers. This is different from our proposed multi-server FL architecture, where the local model updates are required to be aggregated on regional servers. However, even though the network of decentralized FL is well-connected, it is not avoidable to reduce the degradation due large communication delay, since the bandwidth of each client should be much less than edge computing.

\section{Conclusion}\label{Sec:Conclusion}
In this paper, we proposed the MS-FedAvg algorithm and presented theoretical analysis on non-iid datasets in general non-convex settings on a multi-server FL architecture with overlapping areas, which can reduce the transmission latency compared to traditional single-server FL. Our theoretical results reveal how the overlapping areas accelerate the convergence of the final global model. In addition, MS-FedAvg algorithm achieves a linear speedup under full/unbiased partial client participation strategies compared to the existing multi-server FL algorithms. To further improve the convergence rate, we develop a biased partial client participation strategy. Both theoretical and empirical results show the degree of bias results in a trade-off between convergence rate and accuracy, and outperforms other existing multi-server FL architectures. Although our work is based on the fundamental theory of traditional FL, it also opens to doors to many new interesting questions in FL studies. For the future work, we plan to investigate how to design the algorithms based on the topology of the multi-server FL architecture, and the consensus control.


\bibliographystyle{IEEEtran}
\bibliography{example_paper}

\end{document}